\documentclass{aa}  
\pdfoutput=1

\usepackage[varg]{txfonts} 
\usepackage{natbib}

\usepackage{threeparttable}
\usepackage{graphicx}
\usepackage{amsmath}    
\usepackage{mathtools}
\usepackage{amssymb}    
\usepackage{hyperref}
\hypersetup{
    colorlinks   = true, 
    urlcolor     = blue, 
    linkcolor    = blue, 
    citecolor   = blue 
}
\usepackage{multirow}
\usepackage{tabularx}
\usepackage[group-separator={,}]{siunitx}
\usepackage{bm}
\usepackage{xcolor}
\usepackage[export]{adjustbox}
\usepackage{url}
\usepackage[switch]{lineno}
\usepackage{CJKutf8}
\usepackage[rightcaption]{sidecap}

\begin{document} 

   \title{Quantifying the detectability of Milky Way satellites with image simulations:\\Case study with KiDS}
   
   \authorrunning{S. Zhang et al.}

\author{Shiyang Zhang \inst{1}
        \and
        Hendrik Hildebrandt\inst{1}
        \and
        Ziang Yan \inst{1}
        \and
        Simon E.T. Smith\inst{2}
        \and
        Massimiliano Gatto\inst{3}
        \and
        Massimo Dall'Ora\inst{3}
        \and 
        Crescenzo Tortora\inst{3}
        \and
        Shun-Sheng Li \inst{1,4}
        \and
        Dominik Els\"{a}sser\inst{5}
}

\institute{
Ruhr University Bochum, Faculty of Physics and Astronomy, Astronomical Institute (AIRUB), German Centre for Cosmological Lensing, 44780 Bochum, Germany\\
\email{shiyang@astro.ruhr-uni-bochum.de}
\and
Department of Physics and Astronomy, University of Victoria, Victoria, BC, V8P 1A1, Canada 
\and
INAF - Osservatorio Astronomico di Capodimonte, via Moiariello 16, 80131, Napoli, Italy
\and
Leiden Observatory, Leiden University, Einsteinweg 55, 2333 CC Leiden, the Netherlands 
\and
TU Dortmund University, Department of Physics 
}

  \abstract
  {Ultra-faint dwarf galaxies, which can be detected as resolved satellite systems of the Milky Way, are critical to our understanding of galaxy formation, evolution, and the nature of dark matter, as they are the oldest, smallest, most metal-poor, and most dark matter-dominated stellar systems known thus far. Quantifying the sensitivity of surveys is essential to investigating their capability and limitations in searching for ultra-faint satellites. In this paper, we present the first study of the image-level observational selection function for Kilo-Degree Survey (KiDS) based on the Synthetic UniveRses For Surveys (\textsc{surfs}) based on KiDS Legacy-like simulations. We generated mock satellites and simulated images that included resolved stellar populations of the mock satellites and the background galaxies, capturing realistic observational effects such as source blending, photometric uncertainties, and star-galaxy separation. We applied the matched-filter method to recover the injected satellites. We derived the observational selection function of the survey in terms of the luminosity, half-light radius, and heliocentric distance of the satellites. Compared to the catalogue-level simulation typically used in previous studies, the image-level simulation provides a more realistic assessment of survey sensitivity, accounting for observational limitations that are neglected in catalogue-level simulations. The image-level simulation shows a detection loss for compact sources with a distance $d \gtrsim 100~\rm kpc$. We argue that this is because compact sources are more likely to be identified as single sources, rather than being resolved during the source extraction process.}
   
\begin{CJK}{UTF8}{gbsn}
\maketitle
\end{CJK}

\section{Introduction}
\label{Sec:Intro}

Dwarf galaxies are the most abundant galaxies in the Universe. Among them, dwarf galaxies with absolute magnitudes fainter than  $M_V = -7.7$  (corresponding to  $L \approx 10^5 L_\odot$ ) are classified as ultra-faint dwarf galaxies \citep[UFDs, see e.g.][for a review]{simon2019faintest}. These galaxies are the oldest, smallest, most metal-poor, and most dark matter-dominated stellar systems known, representing the lowest limit of the galaxy luminosity function. They play a significant role in testing and developing fundamental models in astronomy. Due to their high dark matter concentrations, UFDs are important targets for searching for gamma-rays from weakly interacting massive particle (WIMP) dark matter annihilation and decay (e.g. \citealp{ackermann2011constraining, lat2015searching, drlica2015search, ahnen2016limits, di2021multimessenger, acciari2022combined, mcdaniel2024legacy}). These galaxies underwent little evolution and have survived to the present day as pristine relics of the early Universe (e.g. \citealp{bovill2009pre, bovill2011fossils, wheeler2015sweating}). The physical properties and star formation processes of UFDs provide critical insights into galaxy formation and evolution (e.g. \citealp{mashchenko2008stellar, agertz2020edge}) and motivate studies of the physical conditions during the reionisation stage (e.g. \citealp{bullock2000reionization, shapiro2004photoevaporation, weisz2014star}) of the Universe.

Interest in UFDs has increased significantly over the past two decades (e.g. \citealp{willman2005new, bechtol2015eight, laevens2015sagittarius, homma2016new, mau2020two, smith2023discovery, gatto2024new}), especially with the emergence of wide-field imaging photometric surveys. However, the detection of these UFDs remains challenging even with modern facilities because of their low surface brightness. Typically, they are too faint to be observed beyond the Local Volume and have only been identified as resolved stellar populations of the Milky Way or nearby galaxies within a few megaparsecs. These satellite galaxies have mostly been detected in wide-field optical imaging surveys through statistical overdensities of individually resolved stars. The number of confirmed Milky Way satellites has grown to approximately 60, with the inclusion of candidate satellites bringing the total to around 100, thanks to the rapid development of wide-field deep photometric surveys. The Sloan Digital Sky Survey (SDSS, \citealp{york2000sloan}), with its digitised wide-area multi-band optical imaging surveys, has greatly improved the search for Milky Way satellites (\citealp{willman2005new, willman2005newb, belokurov2006faint, belokurov2007cats, belokurov2008leo, belokurov2009discovery, belokurov2010big, grillmair2006detection, grillmair2009four, sakamoto2006discovery, zucker2006curious, zucker2006new, irwin2007discovery, walsh2007pair, kim2015hero}). More recently, searches from the Dark Energy Survey (DES, \citealp{bechtol2015eight, drlica2015eight, kim2015horologium, koposov2015beasts, luque2016digging}), Pan-STARRS (\citealp{laevens2015sagittarius, laevens2015new}), DECam Local Volume Exploration Survey (DELVE, \citealp{mau2020two,cerny2021discovery, cerny2023six}), ATLAS (\citealp{torrealba2016feeble, torrealba2016survey}), HSC-SSP (\citealp{homma2016new, homma2018searches, homma2019bootes, homma2024final}), UNIONS (\citealp{smith2023discovery, smith2024discovery}), and KiDS (\citealp{gatto2024new}) have significantly advanced the search for Milky Way satellites.

Despite significant progress in recent years, many faint and distant satellites still remain undiscovered, which is closely tied to the `missing satellites problem' \citep{klypin1999where, moore1999dark}. It is a classical tension wherein the standard $\Lambda$CDM model predicts many more dark matter halos around Milky Way mass host halos in dark matter-only simulations than observed galaxies. This tension has been largely alleviated \citep{Sales2022} by implementing baryonic processes that suppress star formation, including radiation from reionisation \citep[e.g.][]{bullock2000reionization, bovill2009pre}, ram pressure stripping \citep[e.g.][]{Grcevich2009, Spekkens2014}, feedback from supernova explosions \citep[e.g.][]{Stinson2007, Sawala2010, Hopkins2018}, and radiation winds from young stars \citep[e.g.][]{Hopkins2020}. More contemporary studies have applied observational selection functions derived from photometric data to high-resolution simulations of the Milky Way environment \citep[e.g.][]{tollerud2008, newton2018, nadler2020} to determine the completeness of the Milky Way satellite population. A separate but related area of interest is determining the lowest mass dark matter halo that is able to harbour a luminous galaxy \citep[e.g.][]{BL2020}, which has also been explored through the combination of dwarf galaxy detection limits and Milky Way-mass simulations \citep[e.g.][]{jethwa2018upper, nadler2024}. These powerful methods are explicitly dependent on the observational selection function, making an accurate assessment of detectability essential.

A comprehensive analysis of UFDs as satellites of the Milky Way requires a thorough evaluation of survey sensitivity. The observational selection function, which shows the detectability of a satellite galaxy as a function of heliocentric distance, size, and luminosity, is widely used to quantify the capability of a survey to detect Milky Way satellites (e.g. \citealp{koposov2008luminosity, walsh2008invisibles, drlica2020milky}). These studies estimate the selection function at the catalogue level by injecting mock satellites into a survey stellar catalogue and applying observational uncertainties, such as photometric errors and detection incompleteness in post-processing. This approach provides an effective way to predict the number of satellites that likely exist but remain undetected. However, it does not fully capture the complexities of the imaging and subsequent data reduction, including the effects of point spread functions (PSF), saturated stars, blending, and background noise. 

In this study, we present the image-level measurement of the observational selection function of KiDS with multi-band images simulated by SKiLLS (\citealp{li2023kids}). KiDS has observed about 1350 square degrees of the sky and covered nine optical and near-infrared bands in conjunction with infrared data from the VISTA Kilo-degree INfrared Galaxy survey (VIKING, \citealp{edge2013vista}). In its fifth data release \citep{wright2024A&A...686A.170W}, KiDS reports $5\sigma$ limiting magnitudes of 24.79 in the $r$-band and 24.96 in the $g$-band, with a median seeing of $0.7''$ in the $r$-band. With good seeing conditions and multi-band coverage, KiDS provides deep and accurate star catalogues, making it a valuable survey for the research of resolved stellar systems in the local volume.

SKiLLS is a suite of multi-band image simulations for weak lensing analysis of KiDS, using customised input galaxy and star mock catalogues, combining cosmological simulations with high-quality imaging observations to achieve realistic galaxy properties, including multi-band photometry, galaxy morphology, and their correlations \citep{li2023kids}. We generated close to ${\sim} 20~000$ resolved mock satellites with a variety of stellar masses, heliocentric distances, sizes, ellipticities, and position angles and embedded them into 100 KiDS-like images. We then applied a matched-filter method  (\citealp{koposov2008luminosity, walsh2008invisibles}) to search for the simulated satellites. The observational selection function was quantified by cross-matching the detection catalogue with the input satellite catalogue. This selection function can be used to estimate the abundance of Milky Way satellites and assess the confidence of a candidate satellite based on its size, absolute magnitude, and heliocentric distance.

This paper is structured as follows. In Sect. \ref{Sec:data}, we introduce the KiDS data and the simulation of KiDS-like multi-band images. Section \ref{Sec:methods} details the methods of satellite simulation and the satellite search algorithm. We present our results on the observational selection function derived from the KiDS survey in Sect. \ref{Sec:Results} and present our conclusions in Sect. \ref{Sec:Conclusions}.

\section{Simulation set-up}
\label{Sec:data}

In this work, we simulated the nine-band KiDS-like images by combining mock galaxies and foreground Milky Way stars using a \textsc{Galsim} package-based \texttt{MultiBand\_ImSim} pipeline~\citep{li2023kids}\footnote{\url{https://github.com/KiDS-WL/MultiBand_ImSim.git}}. KiDS \citep{dejong2013ExA....35...25D} is an optical wide-field imaging survey conducted with the OmegaCAM camera on the VLT Survey Telescope (VST, \citealp{capaccioli2011vlt}) to study weak gravitational lensing and the large-scale structure of the Universe. It covers $\sim$$1350 \rm ~deg^2$, with its footprint divided into two patches: KiDS-N, along the celestial equator, and KiDS-S, in the southern sky. In conjunction with the VISTA Kilo-degree Infrared Galaxy Survey (VIKING), it provides photometry in nine optical and near-infrared bands: $ugriZYJHK_{\rm s}$ \citep{kuijken2019fourth}, making it suitable for the photometric redshift estimation and redshift distribution calibration.

We used galaxy and star catalogues as inputs to the pipeline to produce KiDS-like images. The input galaxy catalogue was drawn from SKiLLS, a mock galaxy catalogue that combines cosmological simulations with high-quality imaging observations to achieve realistic galaxy properties, including multi-band photometry, galaxy morphology, and their correlations \citep{li2023kids}. The backbone cosmological simulations are the \textsc{surfs} $N$-body simulations \citep{Elahi2018MNRAS}, with galaxy properties generated using the open-source semi-analytic model \textsc{Shark} \citep{Lagos2018MNRAS}. The observed galaxy morphology was modelled using a S\'ersic profile \citep{sersic1963BAAA641S}. The structural parameters involved were derived from high-quality imaging observations obtained with the Advanced Camera for Surveys instrument on the \textit{Hubble} Space Telescope~\citep{Griffith2012ApJS}. For details on the learning algorithm and validation, we refer to \citet{li2023kids}. We used all the galaxies from the input catalogue and placed them randomly within each image, ensuring a sufficient separation between galaxies to avoid overlap or clustering effects. The input star catalogue included satellite member stars with nine-band synthetic magnitudes generated from the PARSEC model \citep{bressan2012parsec} and foreground stars derived from the \textsc{Trilegal} model \citep{girardi2005star}. Each simulated image covers $108~ \rm deg^2$ (108 tiles), with observational conditions, including nine-band noise and PSF information, selected from the full KiDS-DR4 \citep{kuijken2019fourth} dataset of 979 tiles to represent the fiducial settings of the KiDS observations. The tiles were stacked using the \textsc{SWarp} software \citep{2010ascl.soft10068B} following the same configurations as in the KiDS pipelines.

For the simulated images, the photometry was measured using the Gaussian Aperture and PSF (\textsc{GAaP}) pipeline \citep{kuijken2015gravitational}. \textsc{GAaP} applies homogenised Gaussian PSFs and elliptical apertures adapted to each galaxy’s shape and size to mimic the KiDS photometric process. It focusses on the high-signal-to-noise-ratio regions of the sources, ensuring consistent and accurate colour measurements across multiple bands; however, this approach might underestimate the total flux, particularly for extended sources.

\section{Methods}
\label{Sec:methods}

\begin{figure*}
  \centering
   \includegraphics[width=\hsize]{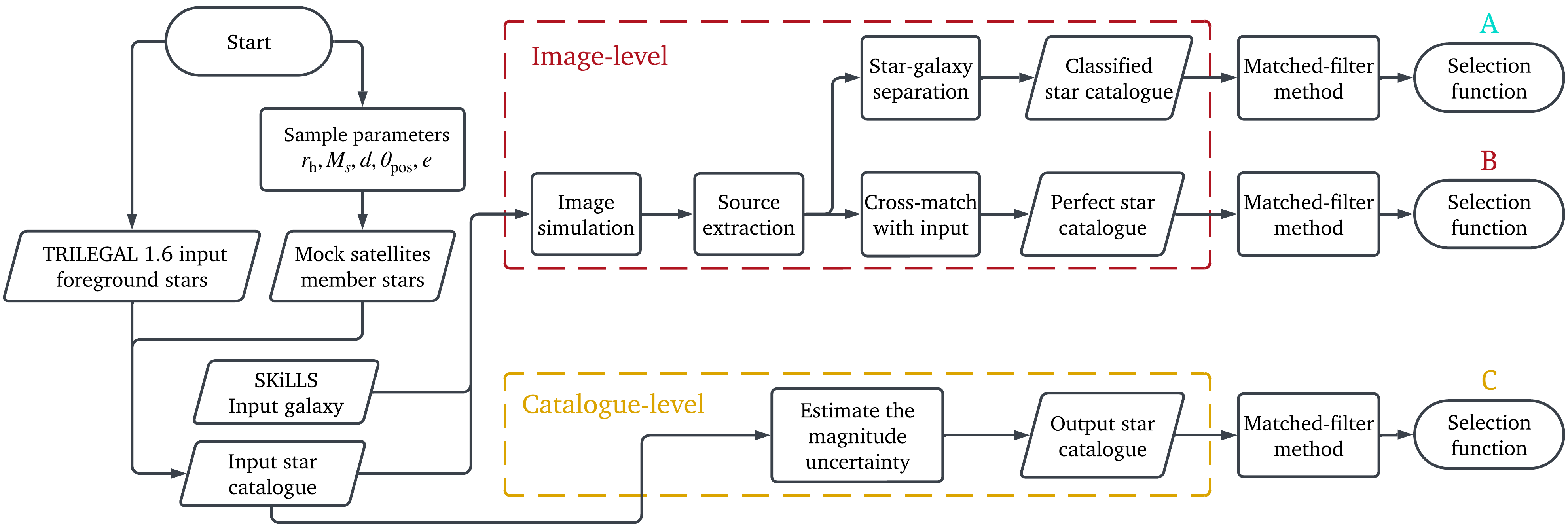}
     \caption{Flowchart summarising the pipeline processes. The output star catalogues are obtained through image-level and catalogue-level simulations. We apply the matched-filter method to recover the injected mock satellites for both catalogues. The observational selection function is constructed by mapping the properties of the satellites to their detectability.}
         \label{Fig:flow}
   \end{figure*}

In this section, we detail the processes of this analysis, including the preparation of the input satellite catalogue, star-galaxy separation, recovery of the mock satellites, and derivation of the observational selection function. We simulated the satellites at both the catalogue-level and image-level, comparing the selection functions of both simulated surveys. The main pipeline processes are shown as a flowchart in Fig. \ref{Fig:flow}.

\subsection{Satellite simulations}
\label{Subsec:SateSim}

\begin{figure*}
  \centering
   \includegraphics[width=17.5cm]{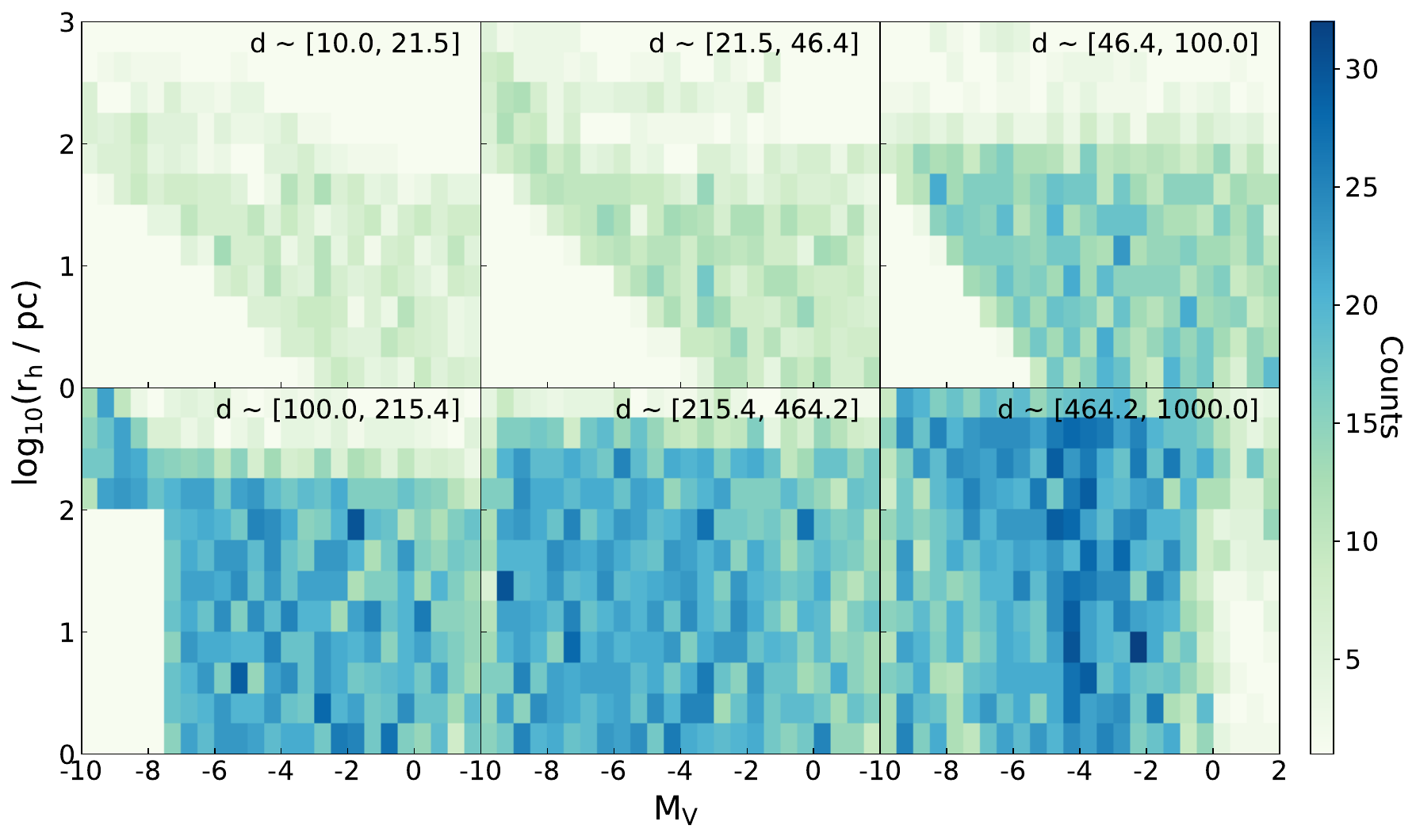}
     \caption{Counts of mock satellites generated in the parameter space. To efficiently use resources for image generation, we did not fully generate mock satellites for very bright and compact satellites, or those with large sizes. For mock satellites with half-light radii exceeding the sizes of the tiles for KiDS-like images, their detection is beyond the capabilities of KiDS and has a lower likelihood of being identified. The non-generated regions will be extrapolated in the observational selection function. The heliocentric distance $d$ is given in kpc.}
         \label{Fig:counts}
   \end{figure*}

We simulated Milky Way satellite galaxies with various properties including stellar mass, heliocentric distance, size, ellipticity, and position angle. To efficiently map the detectability of the satellite searches, we focussed on a three-dimensional parameter space defined by heliocentric distance, half-light radius, and $V$-band absolute magnitude. The parameter space is divided into logarithmically spaced grids: 6 bins for heliocentric distances, $d$; 12 bins for half-light radii, $r_{\rm h}$; and 24 bins for $V$-band absolute magnitudes, $M_{V}$. To optimise the use of computational resources, we generated 5 satellites per grid cell for the first three distance bins and 20 per grid cell for the last three bins. The bright and small satellites were not fully simulated, and only mock satellites with at least 3 detectable member stars were retained, resulting in a total of $19~647$ mock satellites, as shown in the satellite counts per grid in Fig. \ref{Fig:counts}. The properties of the satellites were randomly sampled as described in Table \ref{tab:1}, to form the mock satellite catalogue.

\begin{table}[]
\caption{Properties of the simulated satellites}
\label{tab:1}
\renewcommand{\arraystretch}{1.2}
\begin{tabular}{lccc}
\hline
\hline  

\multicolumn{1}{c}{Parameter} & Range            & Unit          & Sampling \\ \hline
Stellar mass ($M_s$)                 & {[}$10, 10^6${]} & $\rm M_\odot$ & log      \\
Heliocentric distance ($d$)         & {[}$10, 10^3${]} & kpc           & log      \\
Half-light radius ($r_{\rm h}$)            & {[}$1, 10^3${]}  & pc            & log      \\
Ellipticity ($e$)                  & {[}0.1, 0.8{]}   & -           & linear   \\
Position angle ($\theta_{\rm pos}$)               & {[}0, 180{]}     & deg           & linear   \\ \hline
Age ($\tau$)                          & 12               & Gyr           & -      \\
Metallicity ($Z$)                   & 0.0001           & -           & -      \\ \hline
\end{tabular}

\end{table}

To generate a single mock satellite with given properties and location, we sampled the member stars of the mock satellite to match the given stellar mass from an old and metal-poor star population with a total mass of $10^6 M_\odot$. The photometry of the stars was simulated using CMD version 3.7\footnote{\url{http://stev.oapd.inaf.it/cgi-bin/cmd}} with stellar isochrones from \citet{bressan2012parsec}, assuming a metallicity of $Z=0.0001$ and an age of $12 \rm ~Gyr$. The initial masses of the satellite member stars follow a canonical two-part power-law initial mass function (IMF), corrected for unresolved binaries (\citealp{2001MNRAS.322..231K, 2002Sci...295...82K}). Apparent magnitudes were obtained by adding the distance modulus to the given heliocentric distance and we excluded stars with $r$-band magnitudes fainter than 26.5, which approximately corresponds to the $1\sigma$ magnitude limit derived from the $5\sigma$ limit $r$-magnitude of 24.79. The remaining stars, which form the member stars of the mock satellite, were spatially distributed according to a Plummer profile \citep{plummer1911problem}, with a specific half-light radius. The satellite was drawn at the given position coordinates with the specified position angle and ellipticity. This analysis assumes a single stellar population, which does not consider potential variations in age and metallicity. Although no significant bias is expected, such variations could affect the brightness and colour distributions of satellite members, introducing a source of systematic uncertainty in the detectability analysis.

\begin{figure*}
  \centering
   \includegraphics[width=\hsize]{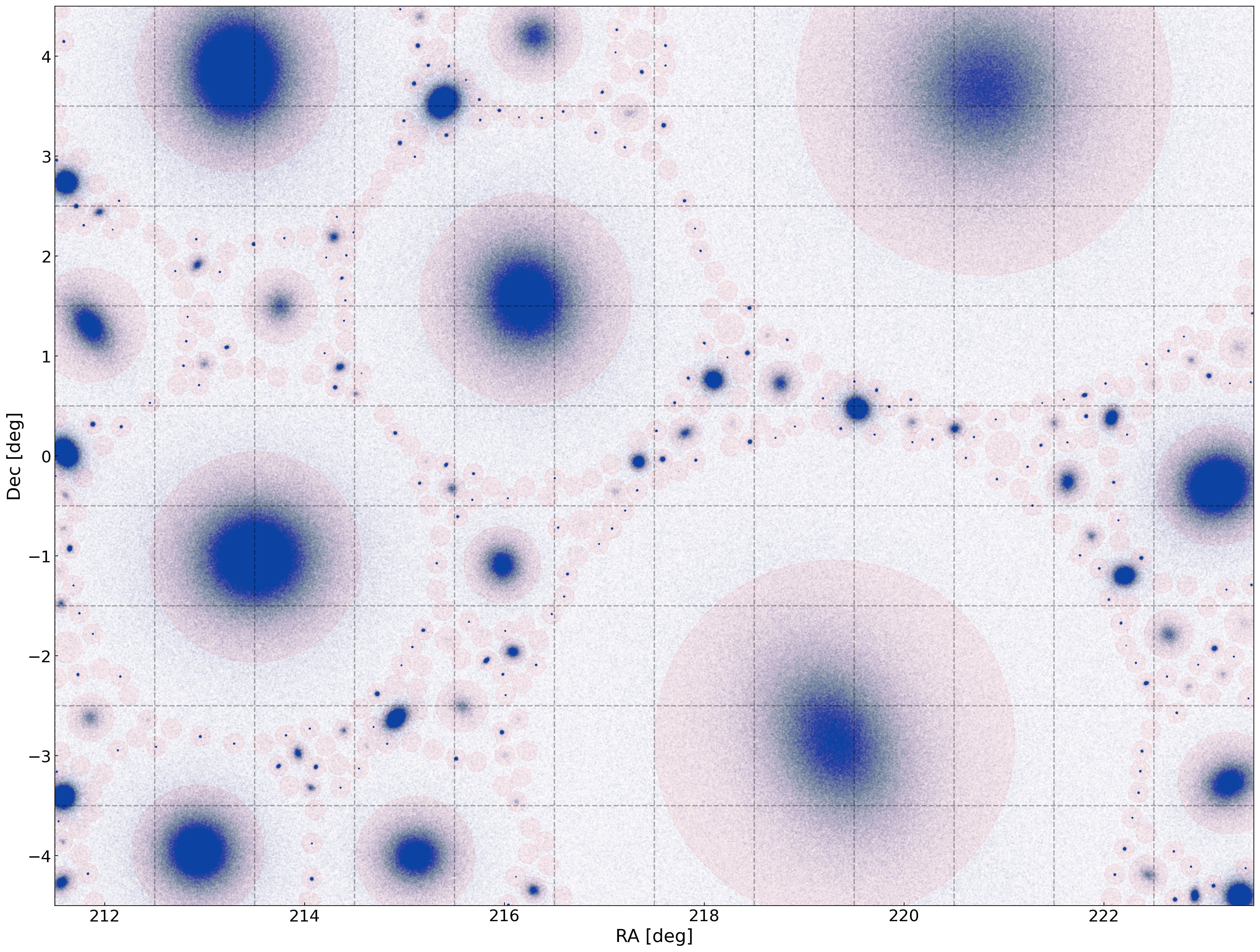}
     \caption{Input star catalogue for one of the $108 \rm ~deg^2$ generated images. Each square block represents a tile in KiDS-like image generation. The red circles indicate the locations of the mock satellites, with the radii representing the smallest separation for each pair of mock satellites with $F=3$ in Eq. (\ref{Eq: 1}). Each blue point represents a star, with the input star catalogue consisting of satellite member stars and foreground stars.}
         \label{Fig:inputs}
   \end{figure*}

We placed the mock satellites on the $108 \rm ~deg^2$ images efficiently, aiming to maximise the number of satellites per image while ensuring that there was adequate spacing between them. We randomly sampled the satellites from the catalogue and placed each mock satellite at a minimum spatial separation of $0.25~\rm deg$. For larger satellites, we determined the minimum separation according to their tangential projection of the half-light radius. The minimal spatial separation can be described as: 
\begin{align}
\label{Eq: 1}
\Delta_{\text{separation}} = \max [F  (r_{\rm new} + r), 0.25~\rm deg] , 
\end{align}
where $r_{\rm new}$ represents the projection of the half-light radius of the next satellite to be placed on the canvas, $r$ represents the projected radius of each already placed satellite, both in units of degrees. Then, $F$ represents a scaling factor for the sum of the sizes, ensuring that the minimum distance between satellites exceeded $F$ times the sum of their radii. To optimise the placement of mock satellites during the generation process, $F$ was dynamically adjusted within a range of 2 to 5, depending on the projected half-light radius of the satellites being placed. Higher values of $F$ were used to ensure wider separations for smaller satellites and the lower values of $F$ were applied to efficiently fit larger satellites into the image and maximise the use of less dense regions. In real observations, the number density of satellites is much lower, making overlaps nearly impossible to detect. This adaptive approach balances between the need  to prevent crowding and the efficient use of the image area, ensuring a proper spacing between satellites. Figure \ref{Fig:inputs} shows an example image, with red circles indicating the positions of each generated satellite. Once the spatial coordinates of the satellites were determined, we placed the satellites with the given properties at these positions. We added foreground stars with their nine-band photometry from \textsc{Trilegal} \citep{girardi2005star}, using the default model from version 1.6 available on the website\footnote{\url{http://stev.oapd.inaf.it/cgi-bin/trilegal}}. We generated the star catalogue for a total field area of $10~ \deg^2$ and simulated each tile image that covers $1 ~\deg^2$ by randomly sampling ten percent of the stars in the catalogue.

\begin{figure*}
  \centering
   \includegraphics[width=17cm]{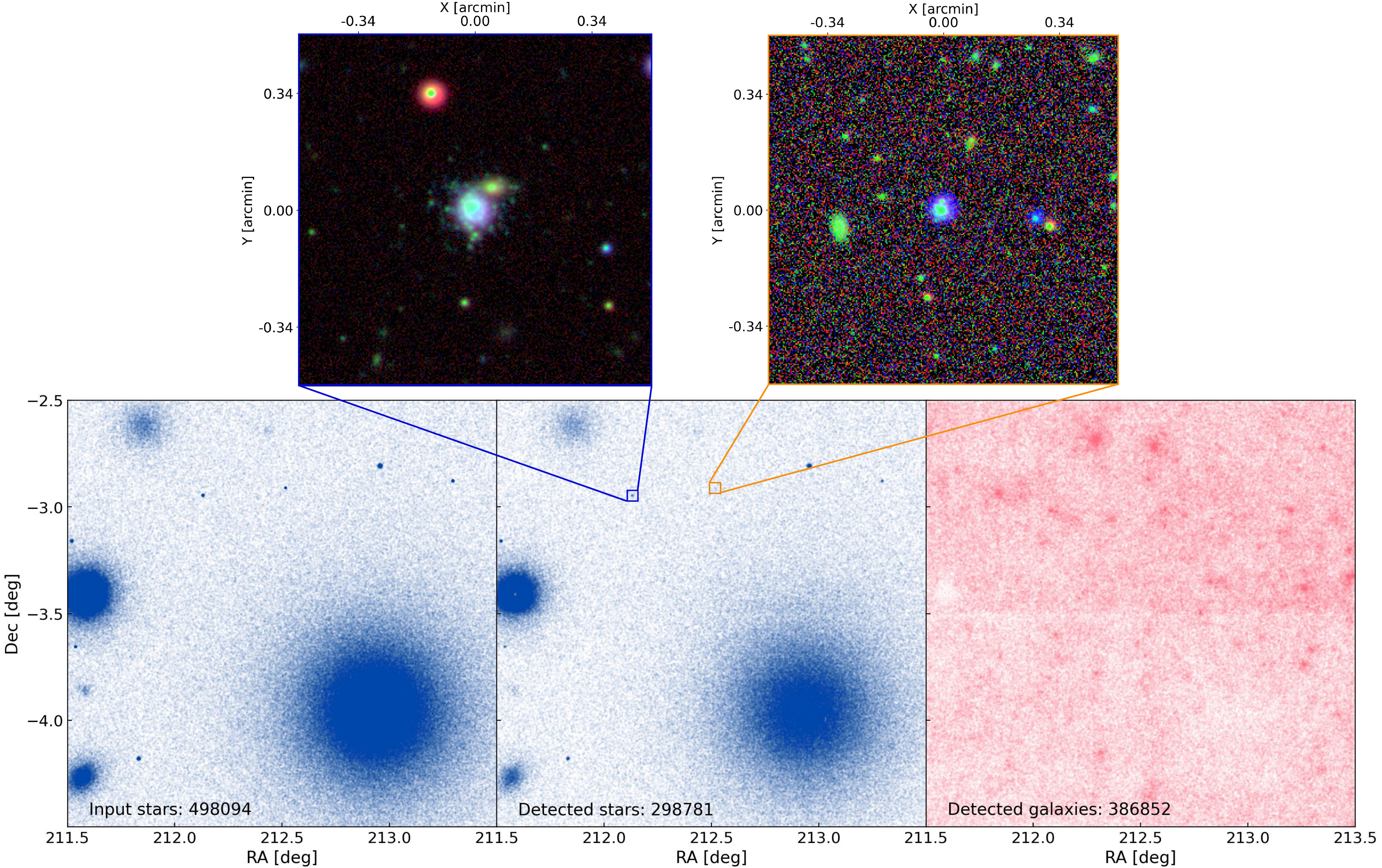}
     \caption{Comparison of the input and detected stars and galaxies for four tiles. The left panel displays the input stars and the middle panel shows the detected stars that are cross-matched with the input catalogue, indicating that only about 60\% of the stars survived the image simulation and source extraction process. The right panel illustrates the detected galaxies identified. Two zoom-in panels show the local environments of two injected satellites. The left panel (blue rectangle) shows a satellite with multiple member stars identified, while the right panel (orange rectangle) shows very few of the member stars are detected, likely due to its compactness and the higher background noise in that region.} 
         \label{Fig:in_out}
   \end{figure*}

For the image-level analysis, we used \texttt{MultiBand\_ImSim} to simulate images and mock the observations according to the input star and galaxy catalogues. The output catalogues of stars and galaxies for each image were obtained by applying \textsc{SExtractor} to extract sources and \textsc{GAaP} to measure photometry. The output stellar catalogues (branch B in the flowchart) were perfectly separated by cross-matching the outputs with the input catalogue from the pipeline. Figure \ref{Fig:in_out} compares the input and output stars and output galaxies in four specific tiles.
Two zoom-in subplots further highlight the local environments of two injected mock satellites. The satellite on the left (blue rectangle) contains multiple resolved member stars that are successfully detected. In contrast, the satellite on the right (orange rectangle) exhibits a very limited member star detection. This injected satellite is more compact and lies in a tile with higher background noise, which likely contributes to its stars blending and make it mis-classified as extended sources.
As simplifying assumptions, we did not include Galactic extinction or spatial variations in stellar density in our simulations. The simulated region spans RA = $[211.5^\circ, 223.5^\circ]$, Dec = $[-4.5^\circ, 4.5^\circ]$, corresponding to moderate Galactic latitudes ($|b| \sim 50^\circ$).
At these Galactic latitudes, foreground extinction is generally low and local stellar density tends to be relatively uniform. 
Based on the KiDS DR4 catalogue, the mean absorption within the simulation region is $0.152 \pm 0.044$ mag for $g$-band and $0.105 \pm 0.031$ mag for $r$-band.
In our simulation, spatial variations in detection depth have been taken into account, with the standard deviation of limiting magnitudes of $0.195$ mag in the $g$-band and $0.188$ mag in the $r$-band. As a result, we expect extinction contributes a smaller level of variation and, thus, has a subdominant effect on the overall sensitivity.
Nonetheless, neglecting these factors may still lead to slight overestimations of sensitivity. Foreground extinction can reduce the brightness of sources, potentially pushing faint objects below the detection threshold, whereas spatial variations in stellar density could increase any source confusion, particularly for faint or diffuse satellites. Therefore, incorporating spatially varying extinction and stellar density into sensitivity modelling would be necessary for analyses requiring higher precision or extending to lower Galactic latitudes, where the extinction and density fluctuations are more significant \citep{drlicaw2021ApJS..256....2D}.

\begin{figure}
  \centering
   \includegraphics[width=\hsize]{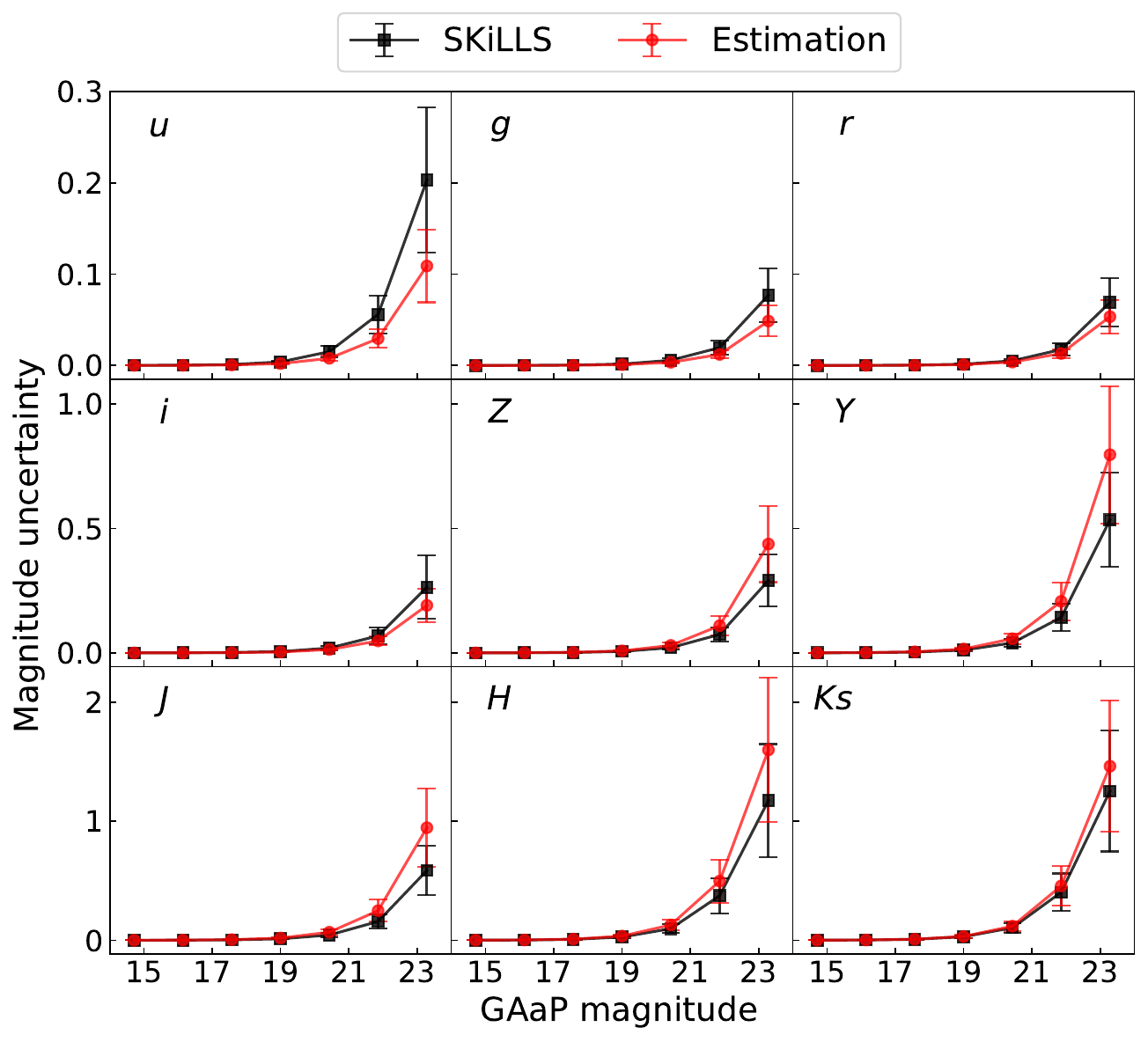}
     \caption{Comparison of the magnitude uncertainties for nine-band photometry with respect to \textsc{GAaP} magnitude ranging from 14 to 24. The black lines represent the \textsc{GAaP} photometric measurements from image-level simulation, and the red lines represent the estimations from Eq. (\ref{Eq: 3}) for a catalogue-level simulation.}
         \label{Fig:magerr}
   \end{figure}

For the catalogue-level analysis (branch C), we used the input star catalogue and estimated the photometric uncertainties based on the depth of KiDS-like images for each band. The flux error, $\Delta f_{i,x}$, for a point source is described as \citep{linke2024euclidkids1000quantifyingimpact}:
\begin{align}
\label{Eq: 2}
5\Delta f_{i,x} = 10^{-0.4(m_{\text{lim},x} - 48.6)} , 
\end{align}
where $m_{\text{lim}, x}$ refers to the $5\sigma$ limiting magnitudes for filter x, taken from \citet{wright2019kids+}. The uncertainty of each band magnitude follows:
\begin{align}
\label{Eq: 3}
m_{\text{err}} = 2.5 \frac{\Delta f_{i,x}}{f_{i,x} \ln{10}}  , 
\end{align}
where $f_{i,x}$ is the flux for each point source, which can be obtained from the given magnitude. We show a comparison of the uncertainties to those from image-level simulations in Fig. \ref{Fig:magerr}.

\subsection{Star-galaxy separation}
\label{Subsec:sg}

We evaluated the impact of the applied star-galaxy separation method on the observational selection function in KiDS surveys (as represented in branch A of the flowchart shown in Fig. \ref{Fig:flow}). We applied a combination of methods, including photometric (Sect. \ref{subsec: bpz}) and machine learning-based approaches (Sect. \ref{subsec: mlp}), to effectively distinguish between stars and galaxies in the catalogue. The image-level simulations are employed to evaluate the point source detection performance of the combined methods. We used two samples, including $216$ tiles with more than $24$ million valid sources. We assessed the point source detection performance using the true positive rate (TPR), false positive rate (FPR) and positive predictive value (PPV), with stars considered as positive and galaxies as negative. The TPR, also referred to as sensitivity or recall, represents the completeness of the star-galaxy separation in detecting stars. It is defined as:
\begin{align}
\label{Eq: tpr}
\rm TPR = \frac{\rm TP}{\rm TP + FN}, 
\end{align}
where TP denotes true positives (correctly identified stars) and FN denotes false negatives (stars misclassified as galaxies). The FPR, also known as contamination, measures the proportion of galaxies incorrectly classified as stars, and it is given by:
\begin{align}
\label{Eq: fpr}
\rm FPR = \frac{\rm FP}{\rm FP + TN}, 
\end{align}
where FP denotes false positives (the galaxies incorrectly classified as stars) and TN denotes true negatives (correctly identified galaxies). The PPV, referred to as purity, indicates the proportion of objects identified as stars that are truly stars, and it can be expressed as follows:
\begin{align}
\label{Eq: ppv}
\rm PPV = \frac{\rm TP}{\rm TP + FP}.
\end{align}
For the purpose of detecting faint dwarf galaxies, it is crucial to make the combined classifier able to identify positive sources with high sensitivity, minimise the number of false detections, and maintain high purity. We applied a magnitude cut at $\texttt{MAG\_AUTO} \leq 25$ to balance completeness and purity in the source catalogue, as objects fainter than this magnitude typically have higher photometric uncertainties in KiDS-like surveys and are more likely to introduce false positives in the classification. An example of the comparison with and without the star-galaxy separator is shown in Fig.~\ref{Fig: combined}.

\subsubsection{Template-fitting with the Bayesian photometric redshift code}
\label{subsec: bpz}

The Bayesian photometric redshift (\textsc{bpz}) code \citep{benitez2000bayesian} is widely used for estimating photometric redshifts from multi-band data based on Bayesian statistics, which uses a library of spectral energy distribution (SED) templates to fit the observed photometric magnitude. Due to its functionality, template-fitting approaches can also be used as star-galaxy classifiers \citep{hildebrandt2012cfhtlens}, particularly at faint magnitudes ($r\gtrsim24$). This is where star-galaxy separation becomes increasingly challenging, as compact galaxies begin to outnumber halo stars \citep{fadely2012ApJ...760...15F}.
 
In our analysis, stars were assumed to have a fixed redshift of $z = 0$, using the \citet{pickles1998stellar} stellar spectral library as their template. For galaxies, the redshift was allowed to vary freely, spanning the range of 0 to 2 with the CWWSB\_capak templates. The \textsc{bpz} code incorporates a spectral energy distribution interpolation of 2 and evaluates redshift steps of d$z=0.05$ to calculate the best-fit estimations for each object along with $\chi^2_{\rm star}$ and $\chi^2_{\rm galaxy}$. The classification is determined by comparing the chi-square values: if a source achieves a smaller $\chi^2_{\rm star}$, indicating a better fit with the star template than the galaxy template, we classified the source as more likely to be a star.

By comparing the difference of the chi-square, the test sample achieves a TPR of 0.6161 and an FPR of 0.0671, indicating that the catalogue is sufficiently clean for further analysis. We note that the classification results are based on simulated data, where the spectral templates used for star generation are closely matched to those in the \textsc{bpz} classifier. In real observations, differences in SEDs, photometric systematics, and variations in galaxy properties may affect the classification performance. Thus, while our method demonstrates the capability of \textsc{bpz} for star-galaxy separation under idealised conditions, its effectiveness in real KiDS data requires further validation.

\subsubsection{Multilayer perceptron classifier}
\label{subsec: mlp}
A multilayer perceptron (MLP) classifier is a neural network-based supervised learning algorithm that can be used to train datasets and classify data into different categories. It consists of multiple layers of interconnected nodes that process and transform input data. 

We customised an MLP star-galaxy classifier using the \textsc{scikit-learn} package\footnote{\url{https://scikit-learn.org/}} \citep{scikit-learn} for KiDS-like catalogues. The input features for the classifier were obtained from extracted source catalogues in the image-level simulations. These catalogues were cross-matched with the input catalogues, allowing each source to be labelled as a star or galaxy. Training features include all columns related to magnitude, size, and flux. Due to the varying depth of each image tile, the performance of the classifier with flux- and radius-related columns shows variation across tiles.

In the image simulation, we mimicked the noise levels of KiDS to approximate realistic survey conditions. To simplify the process, we used the same combination of noise levels across all samples. However, within each sample (108 tiles), the noise levels vary between tiles. As a result, specific tiles across all samples share the same background noise and seeing conditions, which could potentially bias the performance of the classifier, making it location dependent. 

Therefore, to minimise this effect, we trained two MLP classifiers. Specifically, MLP1 has the columns: \{\texttt{MAG\_AUTO, MAGERR\_AUTO, MAG\_GAAP\_0p7\_x, MAG\_GAAP\_1p0\_x, MAGERR\_GAAP\_0p7\_x, MAGERR\_GAAP\_1p0\_x, MAG\_GAAP\_x, MAGERR\_GAAP\_x}\}, where the keyword \texttt{x} represents the different bands and the names containing \texttt{\_0p7} and \texttt{\_1p0} represent the photometry  where the minimum aperture size considered by \textsc{GAaP} is set to $0.7''$ and $1.0''$, respectively; while MLP2 uses the columns \{\texttt{MAG\_AUTO, MAGERR\_AUTO, FLUX\_AUTO, FLUXERR\_AUTO, FLUX\_RADIUS}\}. Although MLP classifiers show relatively low TPR for star-galaxy separation, the high accuracy and effectiveness in identifying true negatives (galaxies) enabled us to confidently remove false detections from the catalogue. 

\subsubsection{Combining strategy}

The combined classifier was constructed considering the performance of both classifiers. Since the classifiers are based on different properties of the stars, the overlap of the candidates from each method increases the confidence in star identification. The \textsc{bpz} template fitting method is effective for initial screening, providing a star sample with a low FPR. However, the clean sample generated by \textsc{bpz} alone does not sufficiently increase the TPR when integrated with the MLP classifiers, which maintain a remarkably low FPR and are suitable for refinement of the selection without having to introduce additional false positives (misclassified galaxies). Therefore, we extended the \textsc{bpz} classified star sample by including sources with $\chi^2_{\rm star} - \chi^2_{\rm galaxy} < 0.5$, thereby forming a broader pool of stars. For MLP classifiers, to ensure robust classification across varying conditions for all tiles, we primarily relied on MLP1 ($p1 > 0.4$), which uses features such as magnitude and photometric errors that are less sensitive to local variations. However, MLP1 alone does not maximise completeness. Therefore, we included high-confidence results from MLP2, applying a stricter threshold ($p2 > 0.8$) to minimise contamination while improving completeness. The union of sources from MLP1 and MLP2 was then intersected with the \textsc{bpz} template fitting results to create the final set of classified sources.

\begin{figure*}
\centering
\includegraphics[width=\hsize]{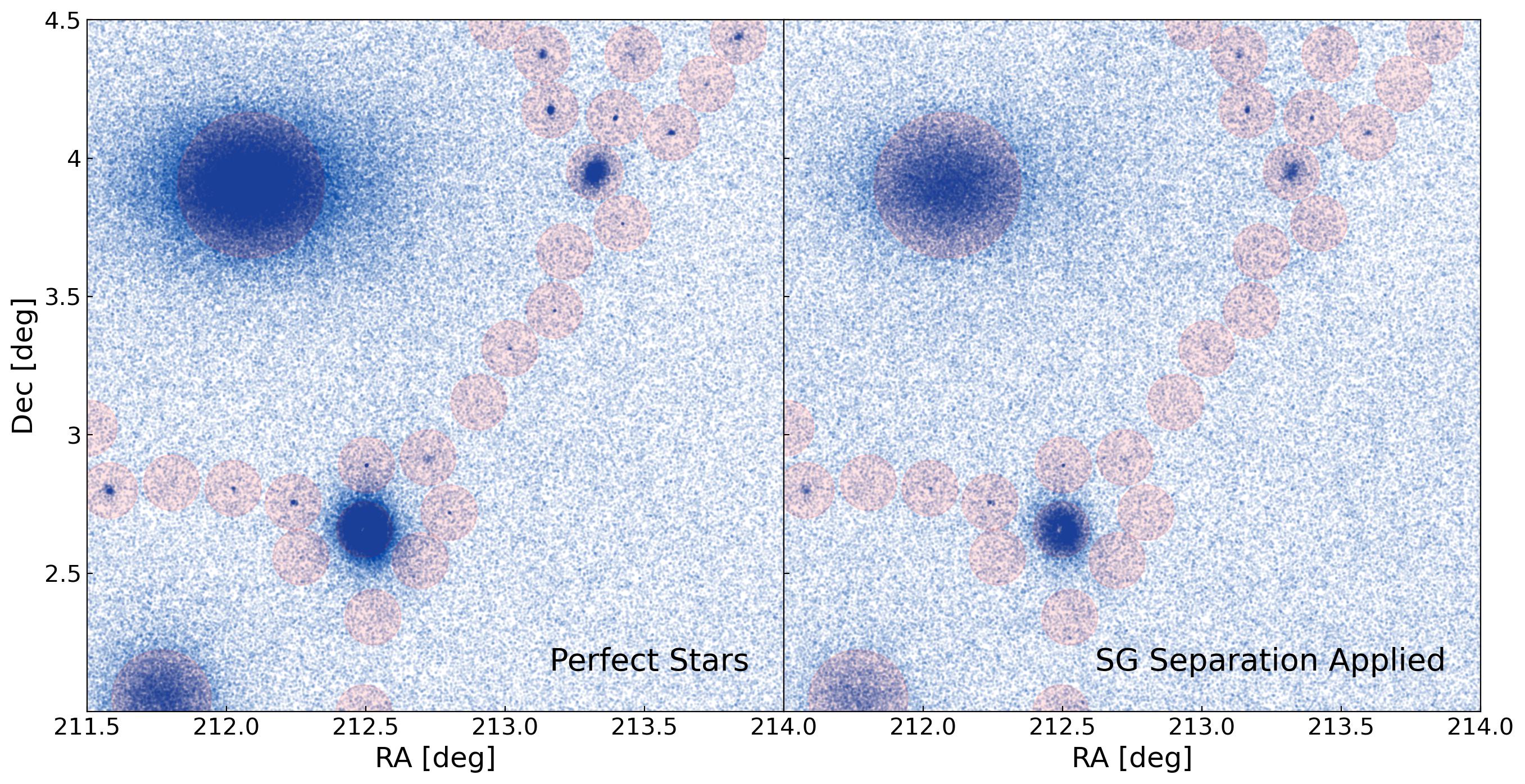}
\caption{Left panel: Detected perfect flag stars are identified by cross-matching the detected sources from \textsc{SExtractor} (branch B) with the input mock satellite catalogue. The red circles represent the locations of the input mock satellites, and the radii of the circles indicate the smallest separation for each pair of mock satellites. Right panel:  Remaining stars after applying star-galaxy separation (branch A), with a \texttt{MAG\_AUTO} cut at 25.}
\label{Fig: combined}
\end{figure*}

\begin{figure}
\centering
\includegraphics[width=\linewidth]{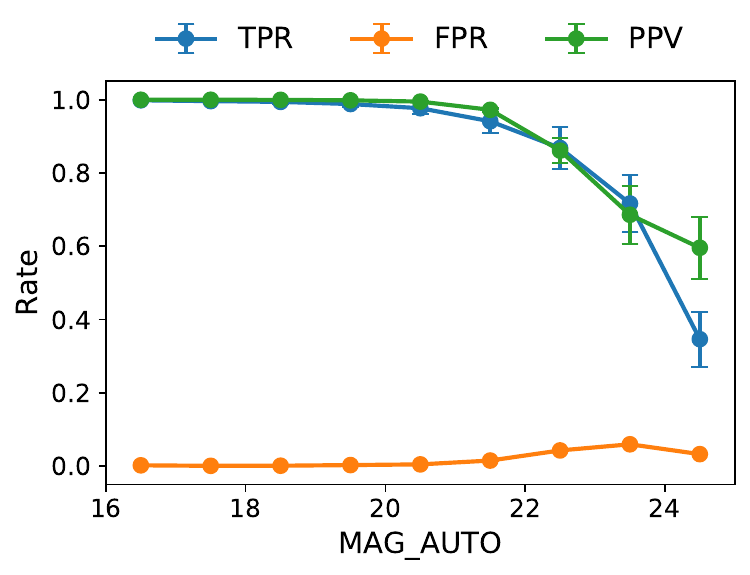}
\caption{True positive rate, false positive rate, and positive predictive value for star-galaxy separation with \texttt{MAG\_AUTO} in the range of [16, 25] for star-galaxy catalogues for all image samples.}
\label{Fig: sg_assess}
\end{figure}

We evaluated the performance of the combined star-galaxy classifier as a function of magnitude, as shown in Fig.~\ref{Fig: sg_assess}. The classifier shows high performance in the brighter region, with a gradual decrease in TPR and PPV as the magnitude increases, while the FPR remains low across the entire magnitude range. This indicates that the classified source catalogue remains clean even in the faint region. The classifier achieves a TPR of $83.13\%$, an FPR of $4.06\%$, and a PPV of $82.65\%$ for \texttt{MAG\_AUTO} $\leq25$ on the test sample. In Fig.~\ref{Fig: combined}, we provide an example of the star catalogue with (right) and without (left) applying the star-galaxy separation in the $2.5^{\circ}\times2.5^{\circ}$ region, where the red circles show the locations of the input mock satellites. From the figure, we observe that the star-galaxy separator functions as expected, without introducing observable contamination, although there is some loss in classifying stars in dense regions.

\subsection{Search algorithm and observational selection function}
\label{Subsec:searching}

We recovered the injected resolved satellites by identifying overdensities of old and metal-poor stars based on their distinct positions in the colour-magnitude diagram (CMD). We adopted a spatially matched-filter method (\citealp{koposov2008luminosity, walsh2008invisibles}) which has been widely applied and proven to be efficient for detecting low-luminosity Milky Way satellites (e.g. \citealp{koposov2015beasts, kim2015hero, kim2015horologium, homma2016new, smith2023discovery, gatto2024new}).

\begin{figure}
  \centering
   \includegraphics[width=\hsize]{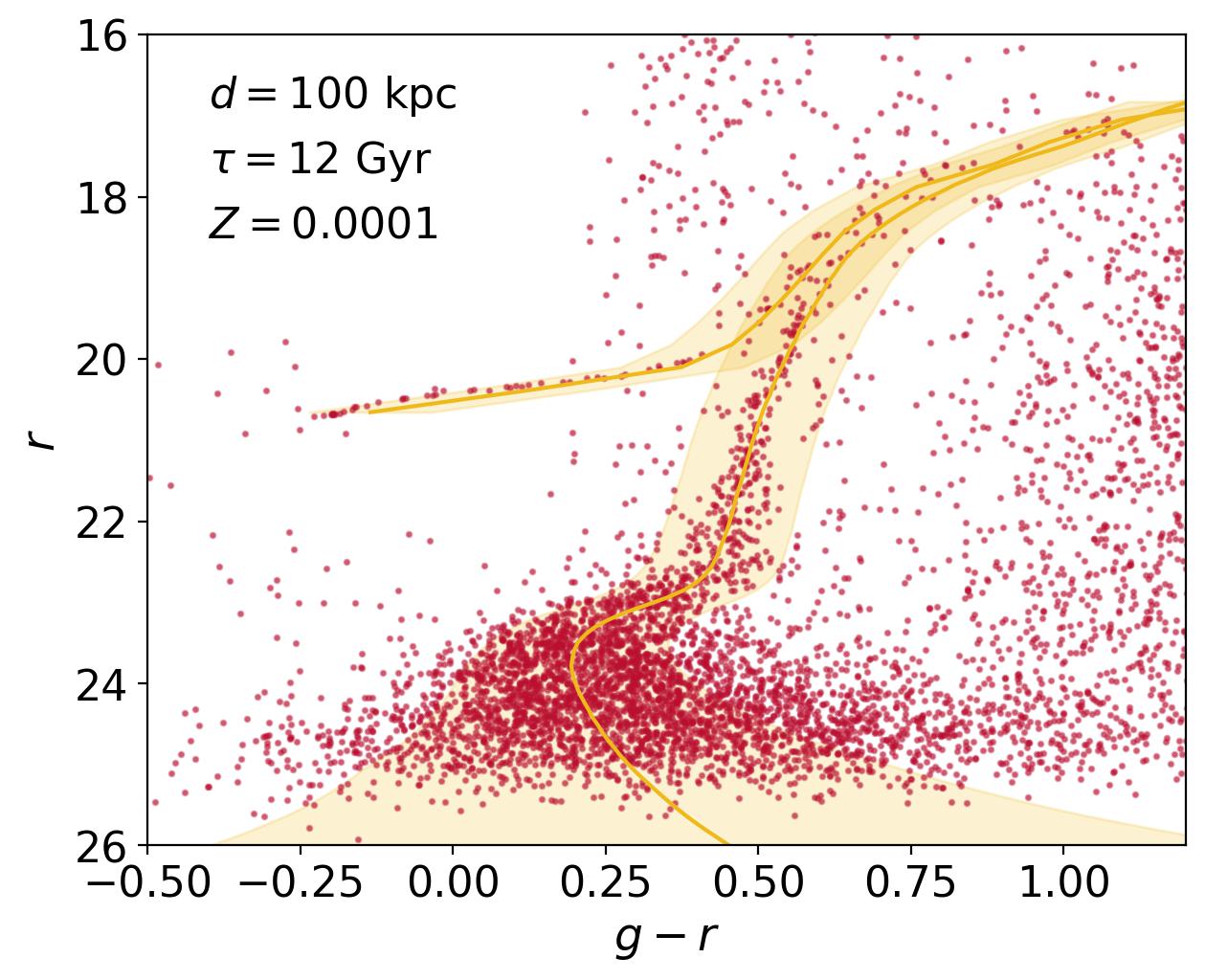}
     \caption{CMD of an example mock satellite galaxy located at a distance of $98.95 ~\rm kpc$, with a half-light radius of $69.45 \rm ~pc$ and an absolute magnitude $M_V = -6.38$ (Chain B). The red points represent point sources selected from the central region of the mock satellite up to 2 times the satellite’s $r_{\rm h}$. These point sources consist of both the mock satellite’s member stars and the Milky Way foreground stars. The yellow region corresponds to the selection criteria derived from Eq. (\ref{Eq: 4}) to isolate the member stars of the satellite. Stars falling within this region are considered as member stars of the mock satellite.}
         \label{Fig:exp_cmd}
   \end{figure}

We selected stars based on their locus in the CMD of $g$ and $g-r$, matching an isochrone with an age of $\tau = 12~\rm Gyr$ and a metallicity of $Z=0.0001$, obtained from PARSEC \citep{bressan2012parsec}. The isochrone was shifted corresponding to logarithmically spaced heliocentric distances $d$ in the range $[10, 1000]$ kpc, divided into ten steps. The criteria for the colour difference between each star and the given isochrone is defined as:
\begin{align}
\label{Eq: 4}
\Delta(g-r) = \sqrt{0.1^2 + \sigma_g^2 + \sigma_r^2}, 
\end{align}
where $\Delta(g-r)$ is the colour difference for each given star and the isochrone at the same magnitude, and $\sigma_{g, r}$ are the photometric uncertainties of the $g$ and $r$ bands. Figure \ref{Fig:exp_cmd} illustrates a representative CMD of a generated mock satellite, including both the mock satellite’s member stars and Milky Way foreground stars, located at a heliocentric distance of $98.95 \rm ~kpc$.

We projected the position of the stars that met the colour-magnitude criteria onto tangential planes and binned them spatially to create stellar number density maps. The projection bin sizes were $0.5\arcmin \times0.5 \arcmin$ and $0.25\arcmin \times0.25 \arcmin$, specified for different projected sizes of the mock satellites injected. The filtered stellar density maps were smoothed with 2D top-hat kernels to estimate the local stellar background, noted as $\rho_{\rm loc}$. The standard deviation of the local stellar background, $\sigma_{\rm loc}$, was calculated as the square root of the variance of the smoothed map:
\begin{align}
\label{Eq: 5}
\sigma_{\rm loc} = \sqrt{\left<\rho_{\rm loc}\right>^2 - \left<\rho_{\rm loc}^2\right>}. 
\end{align}

We also convolved the filtered stellar density maps with 2D Gaussian kernels of different sizes to obtain the smoothed density maps, $\rho_{\rm sm}$, according to the projected sizes of the mock satellites. For stellar density maps, the projection sizes were set to $0.5 \arcmin \times 0.5\arcmin$ and 2D top-hat kernels were chosen from $\{15\arcmin, 20\arcmin, 30\arcmin\} $, and the Gaussian kernel sizes were chosen from $\{1.2\arcmin, 2.4\arcmin, 4.8\arcmin\}$. For projection sizes with $0.25 \arcmin \times 0.25\arcmin$, we adopted $5\arcmin$ and $0.5\arcmin$ for the local stellar backgrounds, $\rho_{\rm loc}$, and smoothed density maps, $\rho_{\rm sm}$, respectively. The significance map was then obtained as the ratio of the statistical overdensities to the local stellar background as:
\begin{align}
\label{Eq: 6}
s = \frac{\rho_{\rm sm} - \rho_{\rm loc}}{\sigma_{\rm loc}}.
\end{align}
We calculated the significance map for a logarithmic range of distances, producing ten maps for each distance by applying various combinations of bin sizes and smoothing kernels. To validate the algorithm, we assessed the capability to identify the known stellar systems within the KiDS footprint and we have successfully recovered satellites such as Sculptor, Fornax, Leo IV, and Leo V. 

We cross-matched the candidate catalogue spatially with the mock satellite catalogue, considering only candidates with a significance greater than the threshold $s \geq 1.5$. The threshold was chosen to balance the detection of satellite candidates while minimising contamination from noise. We found that lowering the threshold led to a significant increase in false detections, with a substantial number of noise fluctuations being misidentified as candidates. We required the centroids of the candidates to be within a radius $\Delta r = \max(0.75\arcmin, r_{\rm h}/F_{\rm m})$, where $r_{\rm h}$ is the projected half-light radius of the mock satellite in arcminutes. Also, $F_{\rm m} = 5$, indicating that the minimal separation for the cross-match with these satellites is within one fifth of the projected radius. 

Following the work from \citet{drlica2020milky}, we presented the observational selection function with heliocentric distance, $d$, absolute $V$-band magnitude, $M_V$, and physical half-light radius, $r_{\rm h}$. These parameters were divided into grids with 6 logarithmically spaced $d$ bins from 10 to 1000 kpc, 24 linearly spaced $M_V$ bins in the range [-10, 2], and 12 logarithmically spaced $r_{\rm h}$ bins in the range [0, 1000] pc. The detectabilities of a total of 19 647 mock satellites were mapped onto the grid map for both image-level and catalogue-level analyses.

\section{Results}
\label{Sec:Results}

\begin{table*}[]
\caption{Parameters for a 50\% detectability limit}
\centering
\label{tab:2}
\renewcommand{\arraystretch}{1.2}
\begin{threeparttable}
\resizebox{\textwidth}{!}{
\begin{tabular}{cc rrrr rrrr rrrr}
\hline
\hline
 &      & \multicolumn{4}{c}{Chain B: image-level\textsuperscript{a}}  & \multicolumn{4}{c}{Chain C: catalogue-level}  & \multicolumn{4}{c}{Chain A: image-level\textsuperscript{b}}       \\
$d$ (kpc)    & $\mu$ (mag)\textsuperscript{c} & $A_{\rm im}$ & $B_{\rm im}$ & $C_{\rm im}$ & $D_{\rm im}$ & $A_{\rm cat}$ & $B_{\rm cat}$ & $C_{\rm cat}$ & $D_{\rm cat}$  & $A_{\rm sg}$ & $B_{\rm sg}$ & $C_{\rm sg}$ & $D_{\rm sg}$ \\ \hline
$[10.00, 21.54]$ & $[15.0, 16.7]$ & $-0.01$ & $-1.60$  & $-0.74$ & $3.60$  & $-0.91$  & $2.91$  & $-8.57$ & $8.68$  & $-0.29$  & $-0.38$   & $-2.02$  & $2.73$ \\
$[21.54, 46.42]$ & $[16.7, 18.3]$ & $-1.31$  & $4.59$   & $-7.66$ & $4.64$  & $-2.73$  & $12.0$  & $-19.9$ & $11.7$  & $-1.17$  & $3.32$    & $-4.46$  & $1.70$ \\
$[46.42, 100.0]$ & $[18.3, 20.0]$ & $-0.95$  & $4.66$   & $-10.8$ & $8.28$  & $-0.85$  & $3.72$  & $-8.55$ & $7.49$  & $-0.28$  & $-0.08$   & $-0.67$  & $0.56$ \\
$[100.0, 215.4]$ & $[20.0, 21.7]$ & $0.24$   & $-1.88$  & $1.44$  & $-0.04$ & $0.22$   & $-1.50$ & $0.15$  & $1.87$  & $0.08$   & $-1.71$   & $2.32$   & $-2.85$  \\
$[215.4, 464.2]$ & $[21.7, 23.3]$ & $0.26$   & $-2.27$  & $3.94$  & $-4.81$ & $-0.15$  & $-0.02$ & $0.06$  & $-1.80$ & $0.36$   & $-3.13$   & $5.69$   & $-6.58$ \\
$[464.2, 1000]$  & $[23.3, 25.0]$ & $0.75$   & $-5.38$  & $10.4$  & $-9.30$ & $-0.26$  & $0.55$  & $-0.39$ & $-2.65$ & $0.97$   & $-7.14$   & $14.9$   & $-14.5$ \\ \hline
\end{tabular}
}
\begin{itemize}
    \item[$^{\rm a}$] Image-level simulation with perfect stars.
    \item[$^{\rm b}$] Image-level simulation with star-galaxy separated stars.
    \item[$^{\rm c}$] $\mu$ represents distance modulus,  defined as $\mu = m - M$.

\end{itemize}

\end{threeparttable}
\end{table*}

By applying the matched-filter method and cross-matching the candidates with the injected mock satellite catalogue, we generated an output catalogue of satellites, with detections marked as 1 and non-detections marked as 0. The outputs are then binned according to their physical properties and the observational selection function is presented using logarithmically spaced heliocentric distance bins ($d$), absolute magnitude ($M_{V}$), and azimuthally averaged projected physical half-light radius ($r_{\rm h}$). The counts of the mock satellites generated in the parameter space are shown in Fig. \ref{Fig:counts}. We applied 2D grid interpolation to fill in the blank grids where no mock satellites had been generated. 

\begin{figure*}
  \centering
   \includegraphics[width=17cm]{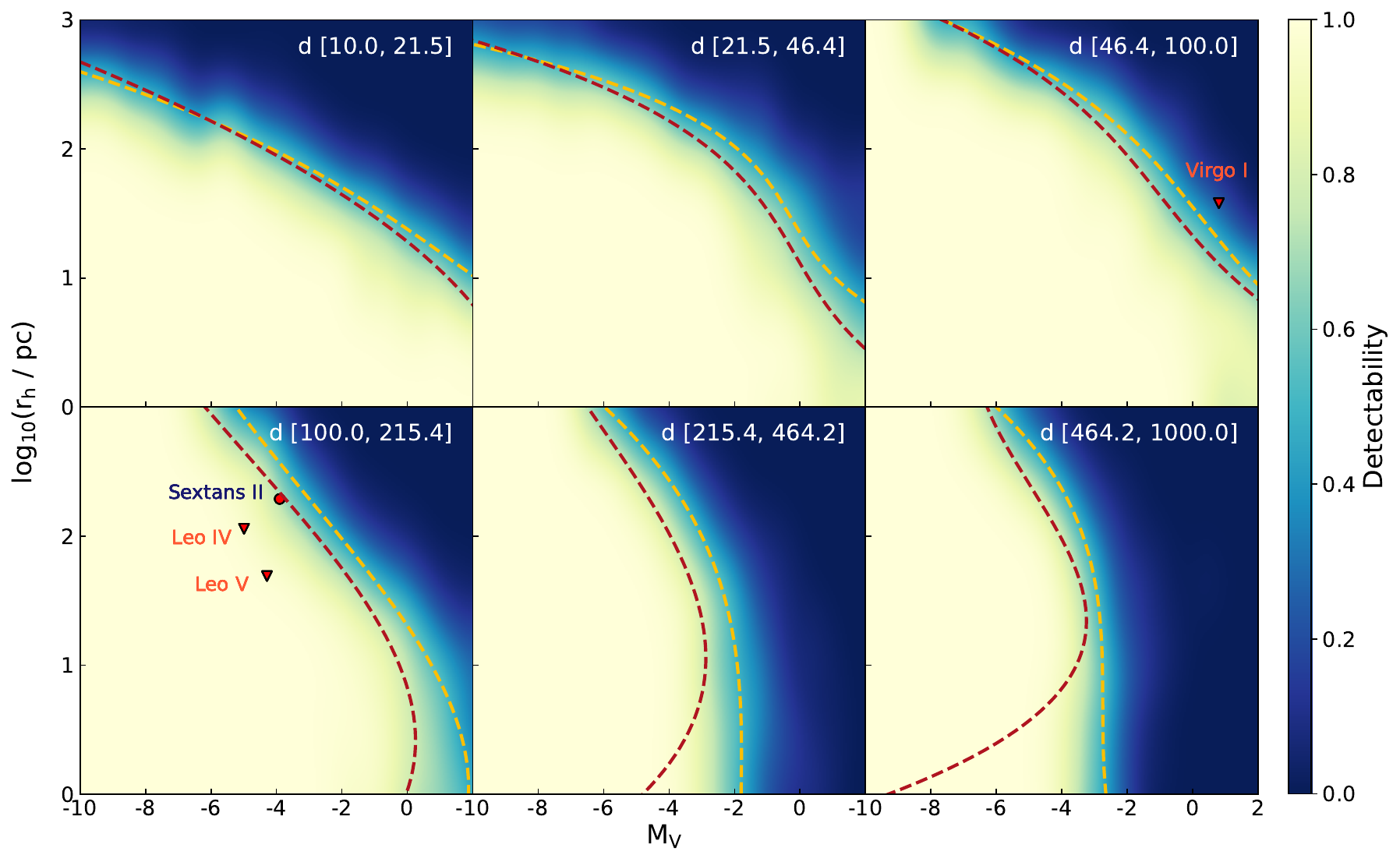}
     \caption{Detection efficiency of searches for satellites for KiDS-like catalogue-level simulation (chain C). The detectability ranges from 0\% (dark blue) to 100\% (bright yellow) as a function of the $V$-band absolute magnitude, physical half-light radius, and heliocentric distance. The 50\% detectability limits fitted by polynomials (n=3) are shown in yellow dashed lines. The detectability limits of image-level simulations are plotted in red dashed lines. The satellites found within KiDS footprint are overplotted in the detection efficiency map.}
         \label{Fig:osf_cata}
   \end{figure*}

\begin{figure*}
  \centering
   \includegraphics[width=17cm]{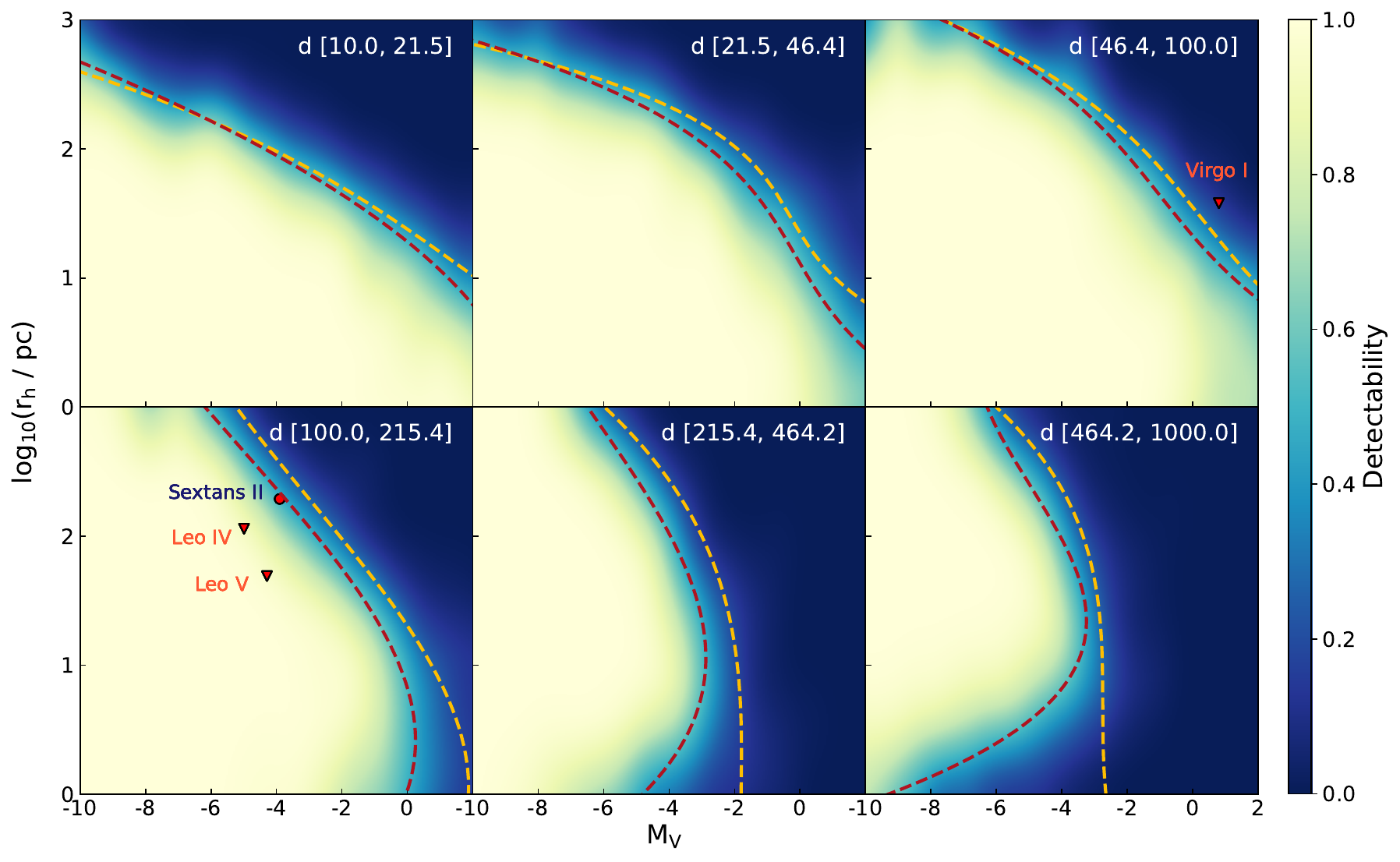}
     \caption{Same as Fig. \ref{Fig:osf_cata}, except for image-level simulation (chain B). The detectability ranges from 0\% (dark blue) to 100\% (bright yellow) similar as shown in Fig. \ref{Fig:osf_cata}. The 50\% detectability limits fitted by polynomials (n=3) are shown in red dashed lines. The detectability limits of catalogue-level simulations are plotted in yellow dashed lines. The satellites found within KiDS footprint are overplotted in the detection efficiency map. }
         \label{Fig:osf_ima}
   \end{figure*}

\begin{figure*}
  \centering
   \includegraphics[width=17cm]{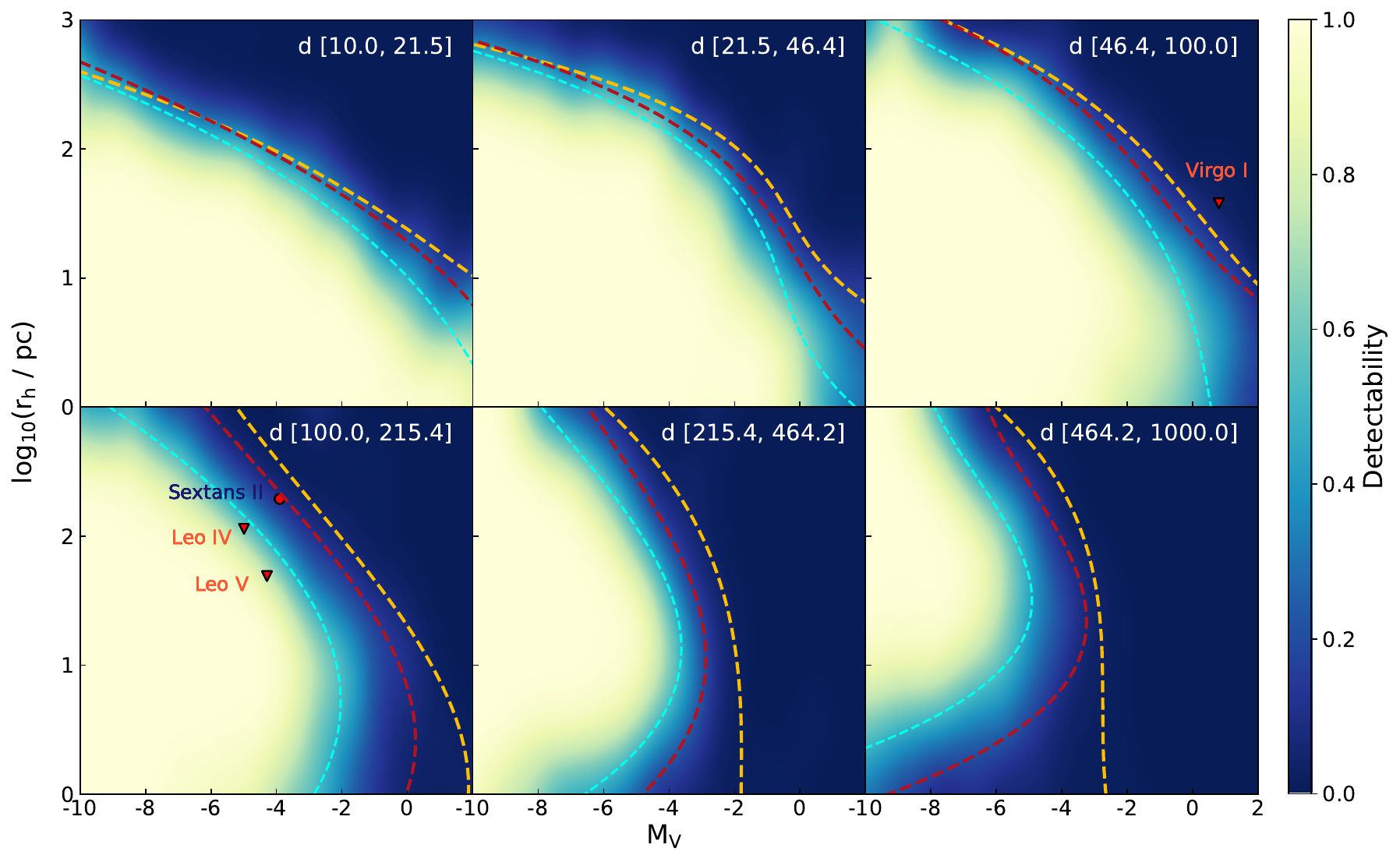}
     \caption{Same as Fig. \ref{Fig:osf_cata}, except with the star-galaxy separator applied in the image-level (chain A). The detectability ranges from 0\% (dark blue) to 100\% (bright yellow) similar as shown in Fig. \ref{Fig:osf_cata}. The 50\% detectability limits fitted by polynomials (n=3) are shown in bright blue dashed lines. The detectability limits of image-level and catalogue-level simulations are plotted in red and yellow dashed lines. The satellites found within in KiDS footprint are overplotted in the detection efficiency map.}
         \label{Fig:osf_ima_sg0}
   \end{figure*}

To fit the 50\% detection probability limit, we first smoothed the detection efficiency map with a Gaussian kernel. The smoothed detectabilities of three different scenarios (corresponding to chains B, C, and A) are shown in Figs. \ref{Fig:osf_cata}-\ref{Fig:osf_ima_sg0}.  We then plotted the 50\% limit $P_{\rm det}(M_{V}, r_{\rm h}) = 0.5$ for each distance bin, and the limits were fitted with polynomials ($n=3$) as:
\begin{align}
\label{Eq: 7}
M_{V} = A_x\rm log_{10}^3 \it r_{\rm h} + B_x\rm log_{10}^2 \it r_{\rm h} + C_x\rm log_{10} \it r_{\rm h} + D_x,
\end{align}
where $r_{\rm h}$ is the half-light radius in each distance bin in units of pc and $M_{V}$ is in units of mag. The distance-dependent parameters $A_x, B_x, C_x, D_x$ were fitted for each distance bin and summarised in Table \ref{tab:2}; the $x$ represents the type of simulation; `im' refers to chain B in Fig. \ref{Fig:flow} using perfect stars, `sg' corresponds to chain A applying star-galaxy separation, and `cat' represents the catalogue-level simulations. We overplotted the $P_{\rm det, 50}$ for both the image- (red) and catalogue-level (yellow) cases as dashed lines.

For the catalogue-level simulations shown in Fig. \ref{Fig:osf_cata}, we find that the shapes of our $P_{\rm det, 50}$ limits are similar to the satellite detectability limits from \citet{drlica2020milky} and \citet{koposov2008luminosity}, indicating that our results are consistent with established findings. The differences between the limits are acceptable because the exact shape of the limits depends on the specific survey and searching algorithm used. Additionally, our results are based on pure simulation data and we employ different distance bins and parameter spaces. 

For the image-level simulations shown in Fig. \ref{Fig:osf_ima}, the detectability of the satellites is lower for the given $M_{V}$ and $r_{\rm h}$ in each distance bin compared to the catalogue-level case. This difference becomes larger as the distance increases. The most noticeable difference in the limits of $P_{\rm det, 50}$ is shown in the lower panel of the detection efficiency maps, where compact satellites are less likely to be detected in the image-level simulation. The turning points of the $P_{\rm det, 50}$ limits occur at about $\rm log_{10}\it r_{\rm h} = \{\rm 0.5, 1, 1.25\} $ in the last three distance bins, corresponding to a projected half-light radius $r_{\rm h, proj}$ of around $0.09~\rm arcmin$.

The loss of compact satellites is probably caused by several effects. We tested insufficient minimum separation during satellite generation as a potential cause by regenerating some of the compact satellites with larger separations, as detailed in Appendix \ref{Sec:compact}. However, the issue of missing compact sources persists. Since this issue only appears in the image-level simulation, the insufficient separation of the satellite is less likely to be the main source of the difference. A potential reason for the loss of compact satellites is that these satellites are so small that they are detected as single sources rather than collections of stars by \textsc{SExtractor}. For example, the input satellite (left panel in Fig.~\ref{Fig:in_out}) with a small size at location $\rm (RA, Dec) = (212.51, -2.91)$ deg is not found in the detected perfect star catalogue (middle panel, in blue rectangle), while larger satellites (e.g. in orange rectangle) can still be detected. Thus, compact satellites with $r_{\rm h, proj} \lesssim 0.09~\rm arcmin$ at heliocentric distances $d \gtrsim 100 \rm ~kpc$ in KiDS-like images have a low probability of being detected using methods based on resolved stars, as these satellites remain unresolved in the detection.

We overplotted the Sextans II discovered with KiDS DR4 \citep{gatto2024new} and independently with HSC-SSP \citep{homma2024final} onto the detection efficiency map for both the image-level and the catalogue-level cases. With a 2D interpolation, the detectability is 48.70\% for the image-level simulation and 71.37\% for the catalogue-level simulation. The detection of Sextans II with KiDS is reliable as it does not exceed the sensitivity limit of the KiDS-like simulation survey. Other satellites that can be found in the KiDS footprint are also plotted on both maps, with their detectability data summarised in Table \ref{tab:3}. 

We also present the smoothed detectability of image-level simulations using the customised star-galaxy separation (chain A) detailed in Sect.~\ref{Subsec:sg}. As shown in Fig.~\ref{Fig:osf_ima_sg0}, the limits of $P_{\rm det, 50}$ are fitted with polynomials ($n=3$) and plotted as bright blue dashed lines on the detection efficiency maps. From the figure, the shapes of the $P_{\rm det, 50}$ limits for the star-galaxy separated scenario are similar to those from the image-level simulation with perfect stars, but are shifted toward the lower left. This shift highlights the significant impact of the star-galaxy separation process on the selection function, where the incompleteness of the star-galaxy classification leads to a loss of faint and extended satellites. The resulting reduction in source significance causes these stellar systems to fall below the detection threshold, making them undetectable with the matched-filter search algorithm. This effect is particularly relevant for real dwarf galaxy searches, where star-galaxy separation is a crucial but challenging step. The observational selection function derived in this study is more restrictive than that from injections at the catalogue-level only, suggesting a more realistic detection capability when accounting for star-galaxy classification effects. Although our star-galaxy separation is only based on simulated data, it highlights the potential application for further refinement and customisation of the star-galaxy separator in future observational studies. It emphasises the importance of incorporating realistic observational effects when assessing survey sensitivity and suggests that improved star-galaxy separation techniques can enhance the detectability of faint satellites.

In addition, since different noise levels and seeing conditions are applied to different tiles, the performance of the star-galaxy separator varies across tiles. The detection efficiency for image-level simulations, with or without the applied star-galaxy separator, reflect an averaged effect on all tiles. In the figure, the detection rate of Sextans II drops to 10.85\% for the star-galaxy separated image-level simulation case, indicating that it is unlikely to be detected under these conditions. However, this value should be considered as a reference only, as Sextans II is located at (RA, Dec) = (156.44, -0.64) deg within the tile \texttt{KIDS\_156.0\_-0.5}. This tile benefits from good seeing conditions, with a full width at half maximum (FWHM) of 0.57 arcsec and a limiting magnitude of 25.02 in the $r$-band. These conditions are expected to contribute to improving source extraction and star-galaxy separation, potentially enhancing the detectability of Sextans II compared to regions with poorer observational conditions.

\begin{table*}[]
\centering
\caption{Detectability of simulated objects resembling known satellites within the KiDS footprint.}
\label{tab:3}
\renewcommand{\arraystretch}{1.2}
\begin{tabular}{ccccccc}
\hline
\hline
Object     & $M_V$ & $r_{\rm h}$ & Distance & Detectability   & Detectability  & Detectability\\
           & (mag) & (pc)        & (kpc)    & cata-level (C) & image-level (B)   & SG separated (A)\\ \hline
Sextans II & -3.90 & 193         & 145      & 71.37 \%        & 48.70\%       & 10.85\% \\
Virgo I    & -0.8 & 38          & 87       & 24.24\%         & 14.66\%        & 3.86\% \\
Leo IV     & -4.99 & 114         & 154      & 96.17\%         & 92.13\%       & 59.40\%\\
Leo V      & -4.29 & 49          & 169      & 97.67\%        & 95.48\%      & 76.11\%\\ \hline
\end{tabular}
\end{table*}

\section{Conclusions and discussions}
\label{Sec:Conclusions}

This study aims to evaluate the sensitivity of KiDS in searching for ultra-faint satellites of the Milky Way, using the first image-level simulation conducted with \texttt{MultiBand\_ImSim} in SKiLLS. To achieve this, we generated $19~647$ satellites arranged in 100 KiDS-like images with various stellar masses, half-light radii, heliocentric distances, ellipticities, and position angles, for each image size of 108 deg$^2$. We established the star catalogue by combining randomly placed foreground stars and satellite member stars sampled from a $Z=0.0001$ and $\tau = 12 ~\rm Gyr$ star pool for each image. For the image-level simulations, we generated the images with the input star and galaxy catalogue and applied \textsc{SExtractor} and \textsc{GAaP} to extract sources and measure their nine-band photometry. For the catalogue-level simulations, the uncertainty of the star photometry was estimated for each star from the input star catalogue. Furthermore, we trained a star-galaxy separation method using a combination of a multilayer perceptron and \textsc{bpz} model fitting. The matched-filter method was applied to recover the injected mock satellites in both image- and catalogue-level simulations, with the image-level simulation including cases with both perfect star detection and star-galaxy separation applied.

We quantified the smoothed observational selection functions for both KiDS-like simulations in terms of $V$-band absolute magnitude, $M_{V}$, half-light radius, $r_{\rm h}$, and the heliocentric distance, $d$, providing selection functions with 50\% detectability limit through analytical fitting with third-order polynomials. Compared with the catalogue-level cases, the image-level cases report a loss of compact sources in the recovery of the satellites; the probable reason is that the compact satellites are more likely to be identified as the single sources during the source extraction process. The results from the catalogue-level selection function qualitatively match the results from \citet{drlica2020milky}, where the catalogue-level dwarf galaxy detection efficiency was assessed for DES and Pan-STARRS. The critical differences between the two studies are the loss of compact satellites and an overall lower detection efficiency as a function of magnitude in the image-level selection function at distances greater than 100\,kpc and a substantial loss in detection efficiency across the board when factoring in star-galaxy separation. The consequence is that our results imply that our current knowledge of the completeness of the Milky Way satellite population is less optimistic than previously thought, as inferred through our method applied to KiDS.

The differences derived from image-level injections provide valuable insights into the `missing satellites problem' and the completeness of the Milky Way satellite census. If the most compact and faint satellites, especially beyond 100 kpc, are systematically missed due to observational selection effects, the apparent dearth of low-mass halos might partially reflect detection limitations rather than a fundamental absence. However, it is important to note that our analysis is based on satellite detection using the matched-filter method, which is optimised for identifying resolved stellar systems. Bright and compact sources may still be detectable through alternative methods. A more comprehensive approach incorporating improved detection techniques and deeper observations will be necessary to fully assess the impact of these selection effects.

Our results also point to possible ways to improve the detectability of ultra-faint satellites in future surveys. The suppression of compact satellite recoveries in image-level simulations suggests that some of these stellar systems might be classified as single sources in the \textsc{SExtractor} pipeline. Adjusting source extraction settings may be helpful in improving the recovery rate of these small mock satellites. Additionally, non-stellar overdensity-based methods, such as machine-learning approaches trained on simulated small and faint dwarf galaxies (e.g. \citealp{Jones2023ApJ...957L...5J}), may provide a more effective means of detecting those sources that are not easily recovered through matched-filter searches. An additional caveat is that all detection efficiency assessments must choose a significance threshold where a mock satellite is deemed to have been detected. However, in all data mining projects that have successfully discovered dwarf galaxies in deep photometric surveys, follow-up imaging and spectroscopy have been necessary to confirm the reality of the faintest candidates, often with a high rate of success. This likely means that all currently derived detection limits are inherently conservative as a function of absolute magnitude. 
While the fully simulated images presented in this study offer a controlled and internally consistent framework for evaluating detection efficiency, they may not fully capture certain observational systematics appear in real survey images, such as PSF variation, image background inhomogeneities, and Galactic extinction. Recent studies (e.g. \citealt{mutlupakdil2021ApJ...918...88M, carlin2024ApJ...977..112C}) adopted synthetic satellite injection into real survey images to empirically evaluate completeness under such effects. These injection-based methods are especially powerful in characterising detection sensitivity within existing observational data, allowing for a detailed assessment of survey completeness around specific galaxies or targeted regions. In contrast, our simulation framework is particularly suited to rapidly and flexibly evaluating detection efficiency for ongoing and planned wide-field surveys. Together, these complementary approaches underscore the importance of carefully balancing observational realism with the capability for systematic exploration when assessing the completeness of satellite surveys.

In the near future, the Legacy Survey of Space and Time at the {\it Vera C. Rubin} Observatory \citep[][]{ivezic2019} will provide groundbreaking photometric depth over much of the extra-galactic sky. In addition, the Euclid Wide Survey from the {\it Euclid} mission \citep[][]{Laureijs2011, ECScaramella2022} will provide the space-based image quality necessary to separate stars and galaxies at extremely faint magnitudes. In tandem, these data sets are expected to enable a huge advance in the detection of extraordinarily faint stellar systems by reducing source blending and improving star-galaxy classification issues that plague current ground-based observations, coupled with the increase in photometric depth. Future analyses should be aimed at integrating image-level effects into survey completeness assessments to ensure a more accurate census of Milky Way satellites and to refine the constraints on dark matter properties derived from their observed population. While this study focuses on UFDs, ultra-diffuse galaxies (UDGs) remain largely absent from our analysis. Their presence and detectability are important issues that  ought to be investigated by future surveys. 

\begin{acknowledgements}
We gratefully acknowledge the anonymous referee for their valuable comments that helped us to improve the manuscript. Special thanks are extended to our colleagues at GCCL for their support and constructive discussions during the preparation of this work. S.Z. acknowledges the support from the Deutsche Forschungsgemeinschaft (DFG) SFB1491.
H. Hildebrandt is supported by a DFG Heisenberg grant (Hi 1495/5-1), the DFG Collaborative Research Center SFB1491, an ERC Consolidator Grant (No. 770935), and the DLR project 50QE2305. ZY acknowledges support from the Max Planck Society and the Alexander von Humboldt Foundation in the framework of the Max Planck-Humboldt Research Award endowed by the Federal Ministry of Education and Research (Germany). M.G. acknowledges the INAF AstroFIt grant 1.05.11. SSL acknowledges funding from the programme "Netzwerke 2021", an initiative of the Ministry of Culture and Science of the State of Northrhine Westphalia and support from the European Research Council (ERC) under the European Union’s Horizon 2020 research and innovation program with Grant agreement No. 101053992. D.E. acknowledges the support from DFG Sonderforschungsbereich 1491 Project F5 and BMBF ErUM-Pro.

\end{acknowledgements}

  \bibliographystyle{aa.bst} 
  \bibliography{aa.bib}

\begin{appendix} 

\section{Detectability and interpolation}
\label{Sec:detect_interp}

\begin{figure}
  \centering
   \includegraphics[width=\hsize]{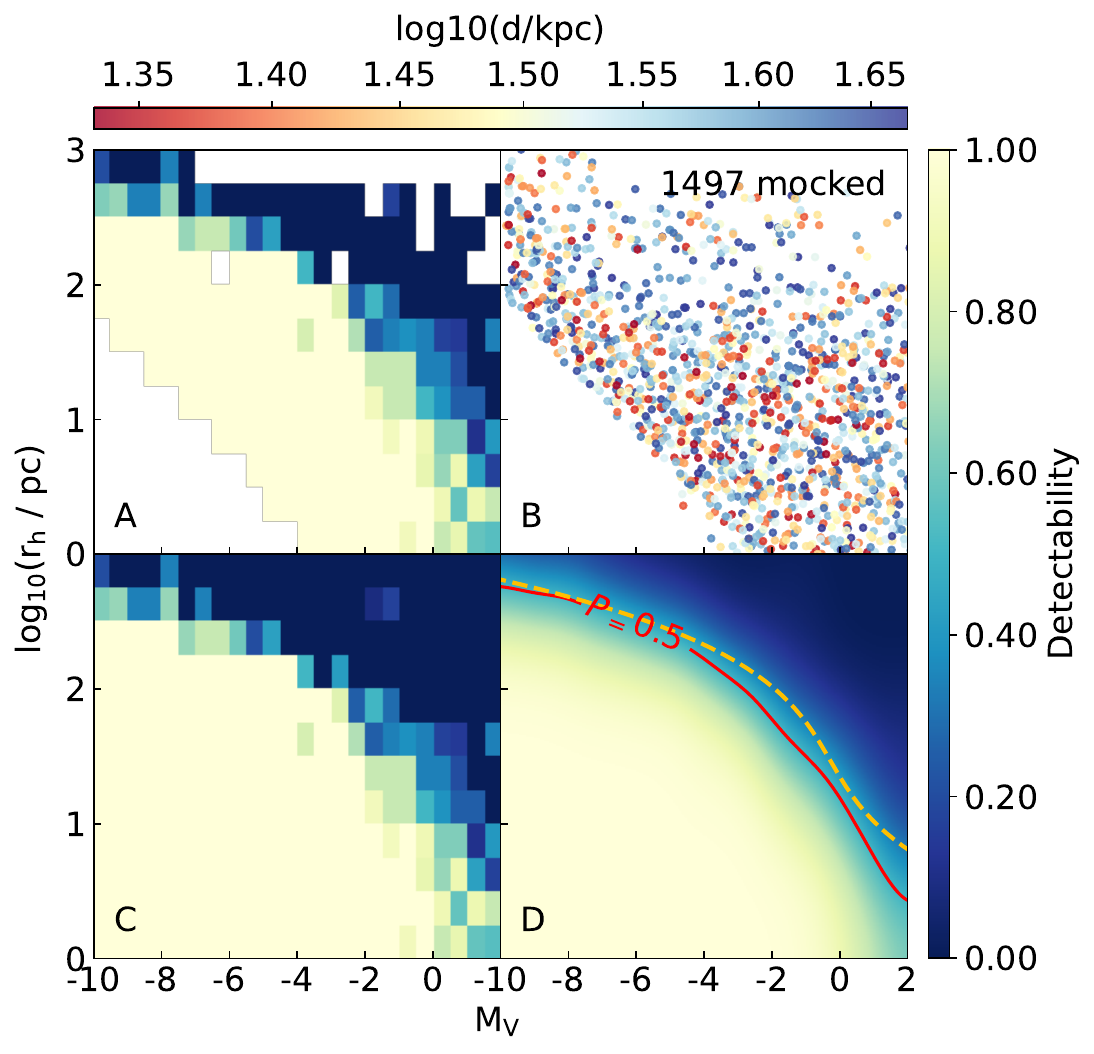}
     \caption{Example of the detectability interpolation and smoothing for the catalogue-level simulation in the second distance bin with $d$ in the range of [21.5, 46.4] kpc. Detection of satellites within this distance range is binned as a function of $M_V$ and $r_{\rm h}$ as shown in subplot A. A total of 1497 mock satellites are plotted as scatters in the $M_V - r_{\rm h}$ map in subplot B. The 2D interpolation is applied to generate the full detection efficiency map, as shown in subplot C. We apply a Gaussian kernel to smooth the map, with the 50\% detectability limit plotted in red in subplot D, and the fitted limit plotted in yellow.}
         \label{Fig:interp}
   \end{figure}

We detail the process of interpolation and smoothing of the satellites detectability in this appendix. We show Fig. \ref{Fig:interp} as an example case of the catalogue-level simulation with a distance range of [21.5, 46.4] kpc. We did not fully generate satellites across the entire parameter space because it would be inefficient for computing resources to simulate many extended satellites in projection size. As shown in subplot B, a total of $1~497$ satellites are generated. For the lower left region, we assume that all satellites generated will be detected, while for the upper region we generate fewer satellites. The colour of each scatter indicates the distance of each mock satellite. We suggest that future results can be further refined by generating more mock satellites around the 50\% detectability limit.
 
\begin{figure*}
  \centering
   \includegraphics[width=17cm]{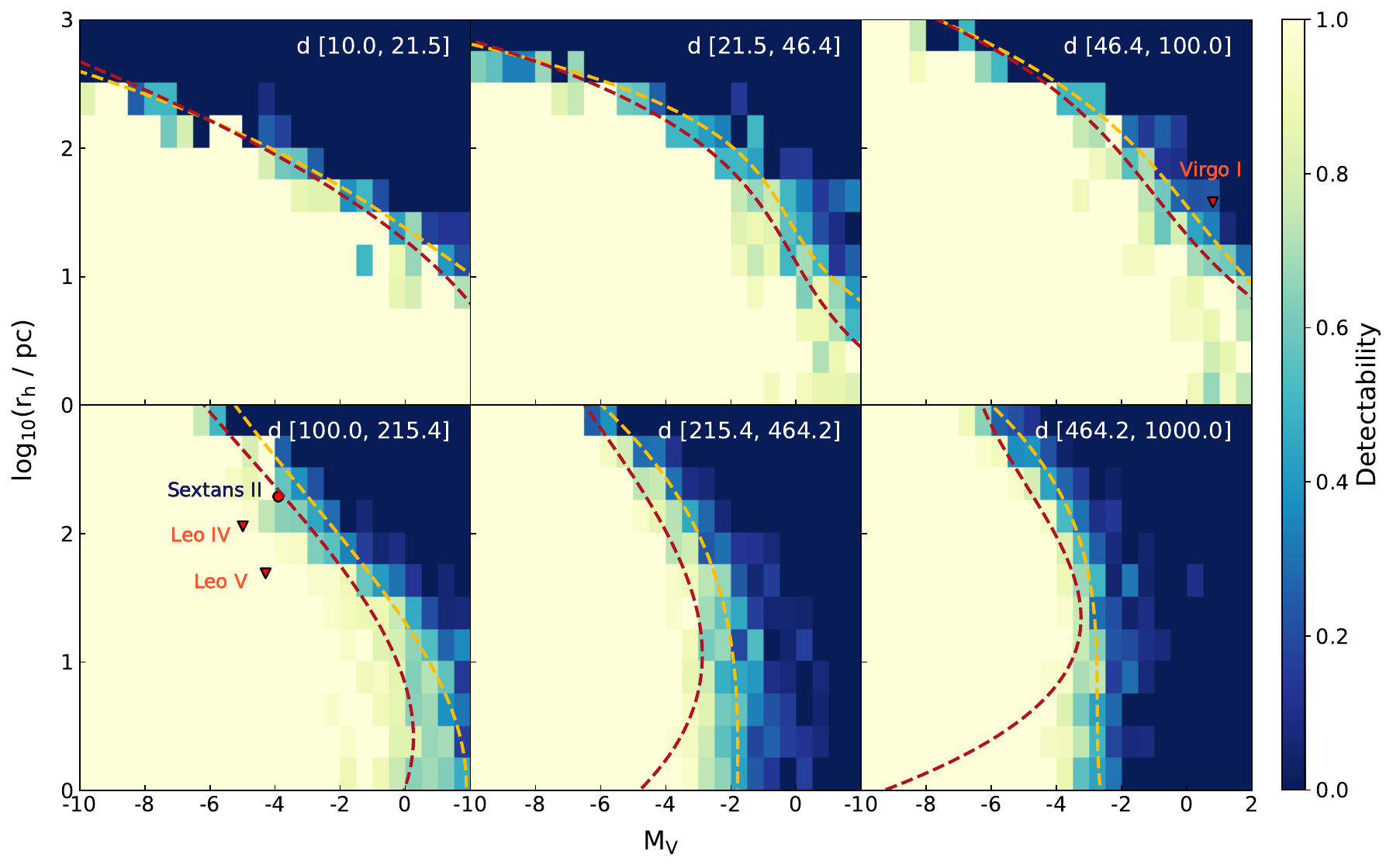}
     \caption{Detection efficiency of satellites searches without smoothing for KiDS-like catalogue-level simulations similar as shown in Fig. \ref{Fig:osf_cata}. The 50\% detectability limits fitted with polynomials (n=3) are shown in yellow dashed lines. The detectability limits from the image-level simulations are plotted in red dashed lines.}
         \label{Fig:osf_cata_bin}
   \end{figure*}

\begin{figure*}
  \centering
   \includegraphics[width=17cm]{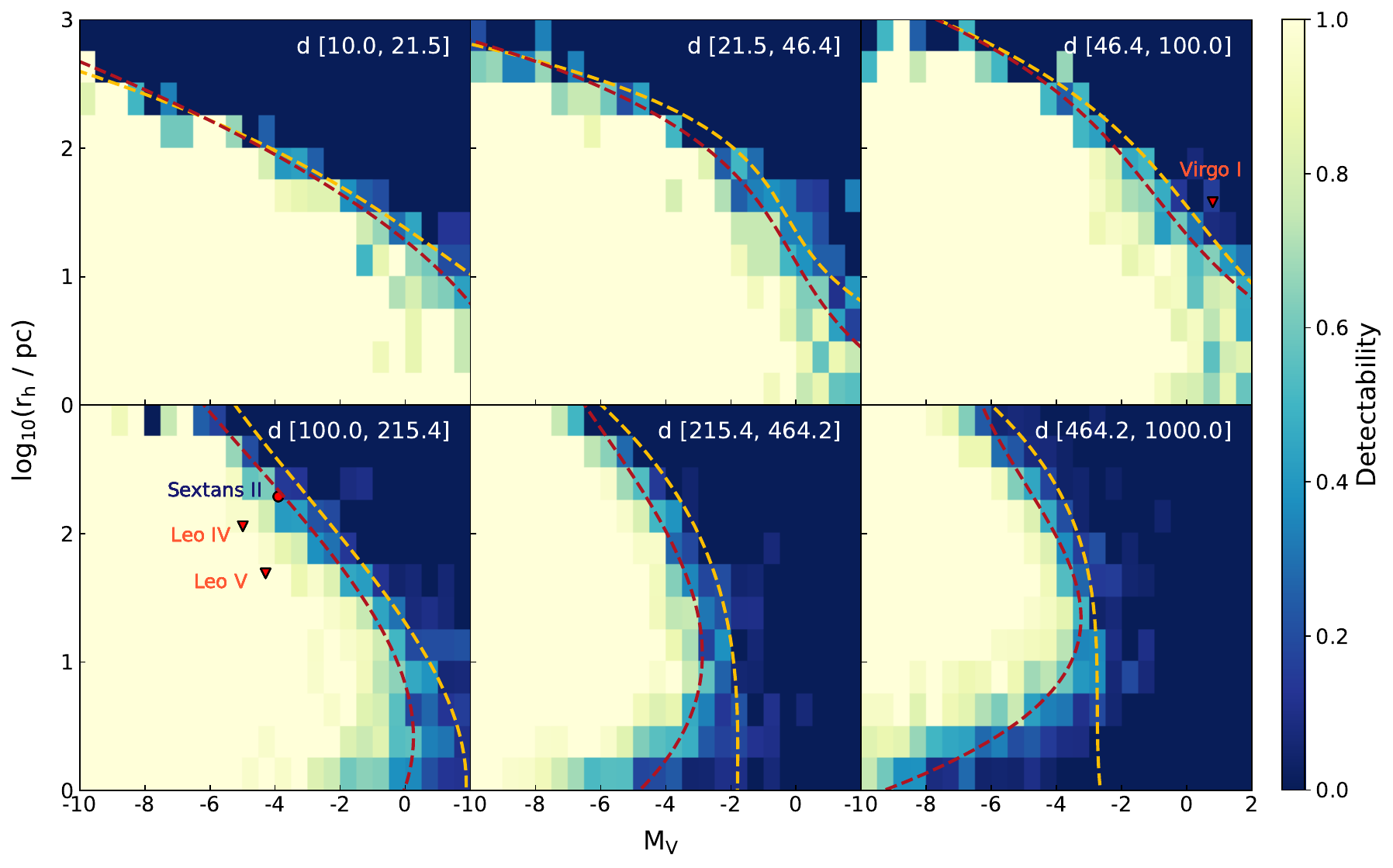}
     \caption{Detection efficiency of satellites searches without smoothing for KiDS-like image-level simulation similar as shown in Fig. \ref{Fig:osf_ima}. The 50\% detectability limits fitted with polynomials (n=3) are shown in red dashed lines. The detectability limits from the catalogue-level simulations are plotted in yellow dashed lines.}
         \label{Fig:osf_ima_bin}
   \end{figure*}

\begin{figure*}
  \centering
   \includegraphics[width=17cm]{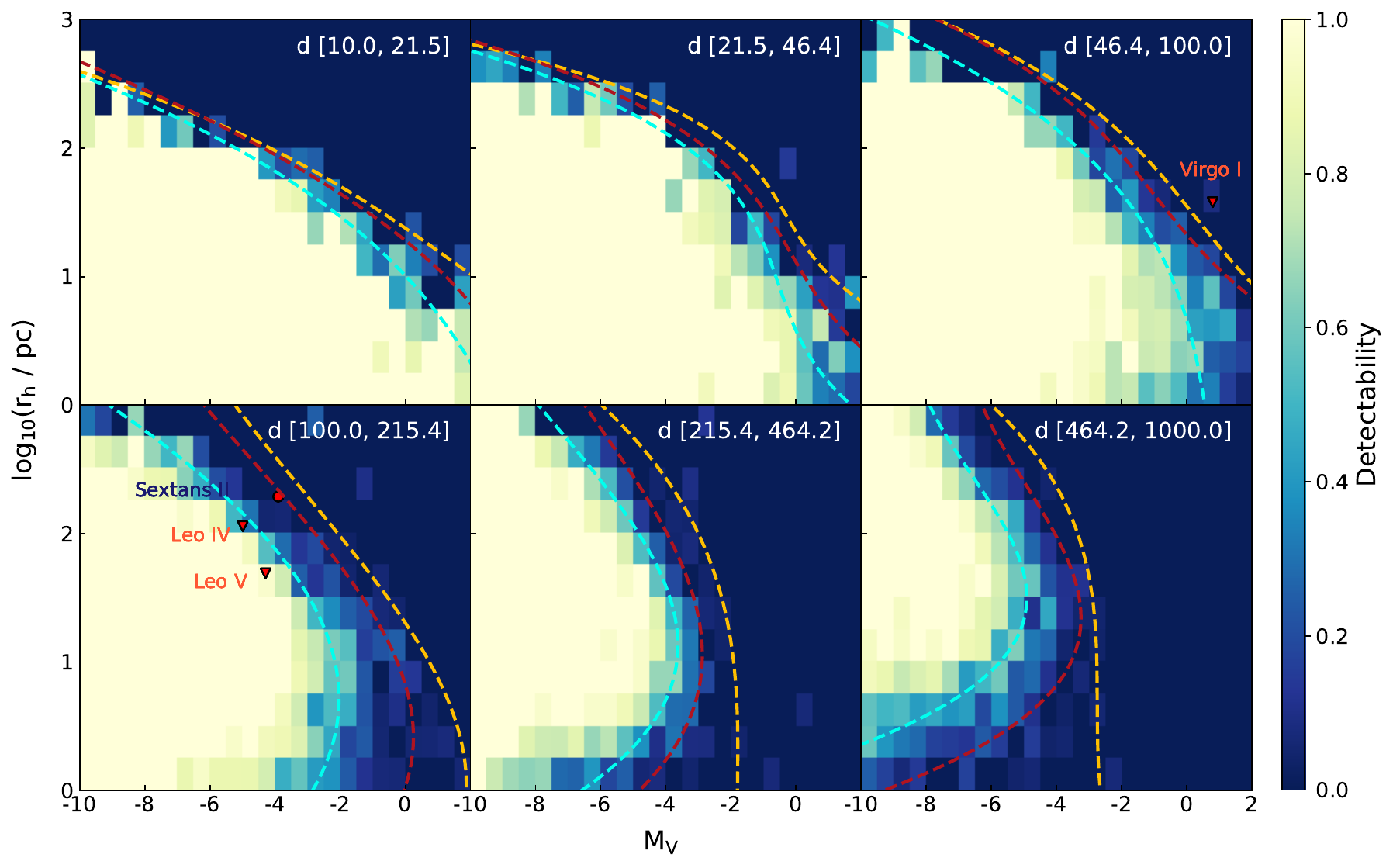}
     \caption{Detection efficiency of satellite searches without smoothing for KiDS-like image-level simulations, with star-galaxy separation applied similarly to Fig. \ref{Fig:osf_ima_sg0}. The 50\% detectability limits, fitted with polynomials (n=3), are shown as bright blue dashed lines. The detectability limits from the image-level and the catalogue-level simulations are plotted as red and yellow dashed lines.}
         \label{Fig:osf_ima_sg0_bin}
   \end{figure*}

The detection catalogue was cross-matched with the input and we binned them according to the absolute magnitude $M_V$ and the half-light radius $r_{\rm h}$, with the original detection efficiency map shown in subplot A. We assume that the detectability of a satellite with parameters corresponding to the top-right grid point of the map results in no detection, namely, $P_{\rm det}(M_V=2, r_{\rm h}=1000 \rm pc) = 0$; whereas satellites with parameters in the lower-left grid point can always be detected, namely, $P_{\rm det}(M_V=-10, r_{\rm h}=1\rm pc) = 1$. We then applied 2D grid interpolation to fill the blanks, as shown in subplot C. The unsmoothed detection efficiency maps for both image-level and catalogue-level simulations are shown in Figs. \ref{Fig:osf_ima_bin} and \ref{Fig:osf_cata_bin} for reference, while Fig.~\ref{Fig:osf_ima_sg0_bin} shows the results of image-level simulation with the star-galaxy separator applied. The Gaussian kernel with $\sigma=1$ is used on the grid map to produce the smoothed detection efficiency map, as shown in subplot D. The 50\% detectability limit $P_{\rm det, 50}$ is overplotted on the map. 

For a fixed distance, fits to the 50\% detectability limit for catalogue-level simulations can be described as an analytical approximation in the form of inverse proportional functions described in \citet{drlica2020milky} or in terms of the limiting absolute magnitude, as in \citet{koposov2008luminosity}. However, for the image-level case, these forms are insufficient to describe the peaks of the limits that appear at $r_{\rm h, proj} \lesssim 0.09~\rm arcmin$ for the heliocentric distance $d>100 ~\rm kpc$. Therefore, we fit the $P_{\rm det, 50}$ limits for both cases with a third-order polynomial, and we plot the fitting limit of this case as a yellow dashed line in subplot D.

\section{Compact sources}
\label{Sec:compact}

In image-level simulations, we find that at distance $d>100 ~\rm kpc$, bright and compact satellites are not fully identified as expected. We also notice that during the image simulation, there are some compact satellites placed close to bright and extended satellites.

To verify whether the proximity of compact satellites to large and bright satellites during image generation can explain the detection loss, we regenerate all the compact satellites that were not detected and had projected half-light radii $r_{\rm h, proj} \lesssim 3.6~\rm arcmin$, placing them with separations described as:
\begin{align}
\label{Eq: 8}
\Delta_{\text{separation}} = \max [F  (r_{\rm new} + r), 0.5 \rm ~deg] , 
\end{align}
similar to the definition in Eq. (\ref{Eq: 1}), $r_{\rm new}$ represents the projection of the half-light radius of the next satellite to be placed on the canvas, $r$ represents the projected radius of each satellite already placed, and they are both in units of degrees. $F$ is the factor for the sum of the projected radii and is selected here as $10$. The minimum separation is set to $0.5 \deg$ for these small satellites. 

\begin{figure*}
  \centering
   \includegraphics[width=\hsize]{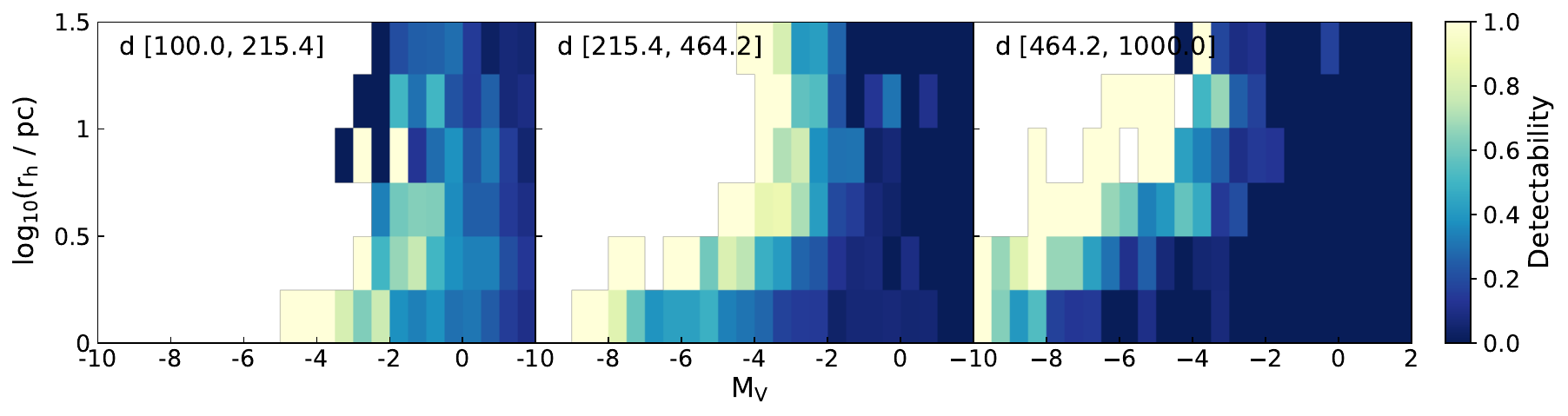}
     \caption{Detection efficiency for searches on compact satellites with $d > 100~ \rm kpc$ in image-level simulations. The figure shows that the loss of detection for compact satellites still persists.}
         \label{Fig:cmp}
   \end{figure*}

After image simulation and detection of the injected satellites, the detectability of compact sources with larger separation is obtained as shown in Fig. \ref{Fig:cmp}. This figure focuses on satellites at distances $d > 100~ \rm kpc$, and the detection for compact satellites is still not complete, suggesting that this issue does not arise from the positioning of small satellites.

\end{appendix}

\end{document}